\title{\textbf{An alternatif test \\ to the two independent samples t test}}
\author{
        \large
        \tt {Rohmatul Fajriyah} \\
        \small
	\tt{Institute of Statistics, TU Graz, Austria} \\
        \small
        \tt{Dept. of Statistics, Univ Islam Indonesia, Jogjakarta, Indonesia} \\
	\href{mailto:fajriyah@student.tugraz.at}{fajriyah@student.tugraz.at}, \href{mailto:rfajriyah@fmipa.uii.ac.id}{rfajriyah@fmipa.uii.ac.id}
}
\date{}

\documentclass[12pt]{article}
\usepackage[left=1.5in, right=1in, top=0.75in, bottom=0.75in, includefoot, headheight=13.6pt]{geometry}
\usepackage[colorlinks=blue]{hyperref}
\usepackage{hhline,multirow,commath,subfig,graphicx,amssymb,amsthm,amstext,amsmath,mathtools,mdwlist,color,float,rotating,rotfloat,caption,xtab,setspace,relsize,exscale,titlesec,soul}
\newtheorem{definition}{Definition}[section]

\usepackage[square, comma, numbers, sort&compress]{natbib}
\usepackage[T1]{fontenc}
\setulcolor{blue}
\setstcolor{red}
\sethlcolor{green}
\numberwithin{figure}{section}
\usepackage{sectsty}

\begin{document}
\maketitle

\begin{abstract}
In this paper we proposed the alternative test to the two independent and normally distributed samples t test based on the cross variance concept. We present the simulation results of the power and the error rate of the special case of the cross variance which is when the variances of the two samples are homogeneous and the t tests. 

The simulation results show that the special case of the cross variance test has the power and the error type I rate equal to the $t$ test. This result suggests that the proposed test could be used as an alternative to detect whether there is difference between means of the two independent normally distributed samples. We give some example comparative case studies of the special case of the cross variance and the t tests.
\end{abstract}

\vskip 1mm
\noindent Key Words: cross variance; error rate type I; power test; t test.
\vskip 3mm

\newpage{}
\pagenumbering{arabic}




\section{Introduction} \label{sec1}

When we have two independent samples and would like to detect whether there is difference between  means of the two populations then we apply the parametric and the non parametric tests. If the samples come from normal distribution, then the $t$ test will be the right choice.

In this paper we will introduce a cross variance concept and present a new approach to detect whether there are differences between the means of the two samples based on the cross variance. 

This paper is organized as follows. In Section~\ref{sec2} we will describe the cross-variance concept and the proposed test, inlcuding the new probability density function. Section~\ref{sec3} will present the simulation results of the power and the error type I of the proposed test in comparison to the $t$ test. Some examples will be provided together with the $t$ test. Finally, section~\ref{sec4} is the summary and remarks.

\section{An introduction to the cross variance concept and the proposed test} \label{sec2}
\subsection{The cross variance concept} \label{sec21}
\begin{definition}
Suppose we have two independent samples, $X_{i}$ and $Y_{j}$; $i=1,2,...,m$ and $j=1,2,...,n$. Their sample mean and variance are denoted by
$\overline{X}, \overline{Y}$ and $V_{x}, V_{y}$. Let 
$$ V_{x}^{*} = \frac{{\sum \limits_{i=1}^{m}(X_{i} -\overline{Y})^{2}}}{m-1} \textrm{  and  } V_{y}^{*} = \frac{ {\sum \limits_{i=1}^{n}(Y_{i} -\overline{X})^{2}}}{n-1}$$
 be the cross-variance for each sample $X$ and $Y$ respectively. The cross-variance sample of groups \textbf{X} and \textbf{Y} is defined as 
\begin{align} \label{eq21}
\textbf{T}&= \frac{V_{x}^{a}+V_{y}^{a}}{2},
\end{align} 
where $\quad V_{x}^{a} = \frac{V_{x}}{V_{x}^{*}}, \quad V_{y}^{a} = \frac{V_{y}}{V_{y}^{*}}.$
\end{definition}
Clearly 
\begin{align}  \label{eq22}
V_{x}^{*} &= \frac{\sum_{i=1}^{m}(X_{i} -\overline{Y})^{2}}{m-1} \nonumber \\
&= V_{x}+\frac{m(\overline{Y}-\overline{X})^2}{m-1},
\end{align}
and
\begin{align}
V_{y}^{*} &=\frac{ \sum_{i=1}^{n}(Y_{i} -\overline{X})^{2}}{n-1} \nonumber \\
&= V_{y}+\frac{n(\overline{Y}-\overline{X})^2}{n-1}. \nonumber
\end{align}
Thus $T= \frac{V_{x}^{a}+V_{y}^{a}}{2}$ can be re-written as
\begin{align} \label{eq23}
T&= \frac{1}{2} \left[\frac{V_{x}}{V_{x} +\frac{m(\overline{Y}-\overline{X})^{2}}{m-1}}+\frac{V_{y}}{V_{y}+\frac{n(\overline{Y}-\overline{X})^{2}}{n-1}}\right] \nonumber \\
&=  \frac{V_{x}}{2V_{x} +2\frac{m(\overline{Y}-\overline{X})^{2}}{m-1}}+\frac{V_{y}}{2V_{y}+2\frac{n(\overline{Y}-\overline{X})^{2}}{n-1}} \nonumber \\
&= Z_{1} + Z_{2}
\end{align}

In what follows, we assume that
\begin{enumerate}
\item{the sample sizes are equal}
\item{${X_{i}}$ and ${Y_{i}}$ are i.i.d. normally distributed with unknown means and known variances $\sigma_{x}^{2}$, $\sigma_{y}^{2}$.}
\end{enumerate}
It follows that
\begin{equation}
\frac{(n-1)V_{x} }{\sigma_{x}^{2}}\sim \chi^{2}_{(n-1)}, \frac{(n-1)V_{y} }{\sigma_{y}^{2}} \sim \chi^{2}_{(n-1)},  \textrm{  and  } \frac{n(\overline{Y}-\overline{X})^{2}}{\sigma_{y}^{2} + \sigma_{x}^{2}} \sim \chi^{2}_{(1)} \nonumber
\end{equation}
Therefore Equation (\ref{eq23}) can be written as follows
\begin{align} \label{eq24}
T&= Z_{1}+Z_{2} \\
&=  \frac{\frac{(n-1)V_{x} }{\sigma_{x}^{2}}}{2\frac{(n-1)V_{x} }{\sigma_{x}^{2}} +2\frac{(\sigma_{x}^{2}+\sigma_{y}^{2})}{\sigma_{x}^{2}}\frac{n}{(\sigma_{x}^{2}+\sigma_{y}^{2})}(\overline{Y}-\overline{X})^{2}}+\frac{\frac{(n-1)V_{y} }{\sigma_{y}^{2}}}{2\frac{(n-1)V_{y} }{\sigma_{y}^{2}} +2\frac{(\sigma_{x}^{2}+\sigma_{y}^{2})}{\sigma_{y}^{2}}\frac{n}{(\sigma_{x}^{2}+\sigma_{y}^{2})}(\overline{Y}-\overline{X})^{2}} \nonumber
\end{align}
where
\begin{align*}
Z_{1}&=\frac{\frac{(n-1)V_{x} }{\sigma_{x}^{2}}}{\frac{2(n-1)V_{x} }{\sigma_{x}^{2}} +\frac{2(\sigma_{x}^{2}+\sigma_{y}^{2})}{\sigma_{x}^{2}}\frac{n}{(\sigma_{x}^{2}+\sigma_{y}^{2})}(\overline{Y}-\overline{X})^{2}}= \frac{U}{2U+2abV},
\end{align*}
and
\begin{align*}
Z_{2}&=\frac{\frac{(n-1)V_{y} }{\sigma_{y}^{2}}}{\frac{2(n-1)V_{y} }{\sigma_{y}^{2}} +\frac{2(\sigma_{x}^{2}+\sigma_{y}^{2})}{\sigma_{y}^{2}}\frac{n}{(\sigma_{x}^{2}+\sigma_{y}^{2})}(\overline{Y}-\overline{X})^{2}}= \frac{S}{2S+2bcV}
\end{align*}
with 
$$
U=\frac{(n-1)V_{x} }{\sigma_{x}^{2}}, \quad S=\frac{(n-1)V_{y} }{\sigma_{y}^{2}}, \quad V=\frac{n(\overline{Y}-\overline{X})^{2}}{(\sigma_{x}^{2}+\sigma_{y}^{2})}
$$ 
and 
$$a=\frac{1}{\sigma_{x}^{2}}, \quad b=(\sigma_{x}^{2}+\sigma_{y}^{2}), \quad c=\frac{1}{\sigma_{y}^{2}}.$$
Hence Equation (\ref{eq24}) can be written as 
\begin{align} \label{eq25}
T &=Z_{1}+Z_{2}=\frac{U}{2U+2abV} + \frac{S}{2S+2bcV}
\end{align}
To compute the distribution of $T$ in Equation (\ref{eq25}), it can be done by considering the fact that
\begin{enumerate}
\item{$U, V$ and $S$ are independent}
\item{$Z_{1}$ and $Z_{2}$ are dependent}
\end{enumerate}
In this paper we will describe the first case, where we consider that $U, V$ and $S$ are independent. 

\subsection{The proposed test} \label{sec22}

Under normality asumption of $X$ and $Y$ then $U, V$ and $S$ are independent, where $V$ is $\chi^{2}_{(1)}$ distributed and $U, S$ are $\chi^{2}_{(n-1)}$ distributed. From the equation (\ref{eq25}), suppose $V=Z_{3}$ from here we get that $U= \frac{2abZ_{1}Z_{3}}{1-2Z_{1}}$ and $S= \frac{2bcZ_{2}Z_{3}}{1-2Z_{2}}$. The jacobian of this transformation is

$|J|=\begin{vmatrix}
  \frac{dU}{dZ_{1}} & \frac{dU}{dZ_{2}} & \frac{dU}{dZ_{3}} \\
  \frac{dS}{dZ_{1}} & \frac{dS}{dZ_{2}} & \frac{dS}{dZ_{3}} \\
 \frac{dV}{dZ_{1}} & \frac{dV}{dZ_{2}} & \frac{dV}{dZ_{3}} 
 \end{vmatrix} =\begin{vmatrix}
  \frac{2abZ_{3}}{\left(1-2Z_{1}\right)^{2}} & 0 & \frac{2abZ_{1}}{\left(1-2Z_{1}\right)}  \\
  0 & \frac{2bcZ_{3}}{\left(1-2Z_{2}\right)^{2}}  & \frac{2bcZ_{2}}{\left(1-2Z_{2}\right) } \\
 0 & 0 & 1
 \end{vmatrix}=\frac{4ab^{2}cZ_{3}^{2}}{\left((1-2Z_{1})(1-2Z_{2})\right)^{2}}$

The joint probability density function of $Z_{1}, Z_{2}, Z_{3}$ is
\begin{align} \label{eq26}
&f_{Z_{1}, Z_{2}, Z_{3}}(z_{1}, z_{2}, z_{3}) \nonumber \\
&=f_{U}\left(u=\frac{2abz_{1}z_{3}}{1-2z_{1}}\right)f_{S}\left(s=\frac{2bcz_{2}z_{3}}{1-2z_{2}} \right)f_{V}(v=z_{3})|J| \nonumber \\
&= \frac{\left(4ab^{2}c \right)^{\frac{n-1}{2}}}{2^{n-\frac{1}{2}}\Gamma \left( \frac{1}{2}\right) \Gamma \left( \frac{n-1}{2}\right)^{2}} \frac{z_{1}^{\frac{n-1}{2}-1} z_{2}^{\frac{n-1}{2}-1} z_{3}^{\frac{2n-1}{2}-1} e^{-z_{3}\left(\frac{1}{2}+\frac{abz_{1}}{1-2z_{1}}+\frac{cbz_{2}}{1-2z_{2}} \right)}}{\left(1-2z_{1} \right)^{\frac{n+1}{2}} \left( 1-2z_{2}\right)^{\frac{n+1}{2}}} \nonumber \\
\end{align}

The joint density function of $Z_{1}, Z_{2}$ is the marginal probability function of $Z_{1}, Z_{2}$ from the equation (\ref{eq26}) above. It is computed as follows:
\begin{align} \label{eq27}
&f_{Z_{1}, Z_{2}}(z_{1}, z_{2}) \nonumber \\
&= \int \limits_{0}^{\infty} f_{Z_{1}, Z_{2}, Z_{3}}(z_{1}, z_{2}, z_{3}) dz_{3} \nonumber \\
&= \frac{\left(4ab^{2}c \right)^{\frac{n-1}{2}}}{2^{n-\frac{1}{2}}\Gamma \left( \frac{1}{2}\right) \Gamma \left( \frac{n-1}{2}\right)^{2}}  \frac{z_{1}^{\frac{n-1}{2}-1} z_{2}^{\frac{n-1}{2}-1} }{\left(1-2z_{1} \right)^{\frac{n+1}{2}} \left( 1-2z_{2}\right)^{\frac{n+1}{2}}} \int \limits_{0}^{\infty} z_{3}^{\frac{2n-1}{2}-1} e^{-z_{3}\left(\frac{1}{2}+\frac{abz_{1}}{1-2z_{1}}+\frac{bcz_{2}}{1-2z_{2}} \right)} dz_{3} \nonumber \\
&= \frac{\left(4ab^{2}c \right)^{\frac{n-1}{2}}\Gamma \left (n-\frac{1}{2} \right) z_{1}^{\frac{n-1}{2}-1} z_{2}^{\frac{n-1}{2}-1} }{2^{n-\frac{1}{2}} \Gamma \left( \frac{1}{2}\right) \Gamma \left( \frac{n-1}{2}\right)^{2} \left(1-2z_{1} \right)^{\frac{n+1}{2}} \left( 1-2z_{2}\right)^{\frac{n+1}{2}} \left(\frac{1}{2}+ \frac{abz_{1}}{(1-2z_{1})}+\frac{bcz_{2}}{(1-2z_{2})}\right)^{n-\frac{1}{2}}} \nonumber \\
\end{align}

Therefore the cumulative distribution function (cdf) of $T \le t$ s computed as follows
\begin{align} \label{eq28}
&F_{T}(t) \nonumber \\
&=P(T \le t) \nonumber \\
&=\int \limits_{-\infty}^{\infty} \int \limits_{-\infty}^{t-z_{1}}f_{Z_{1},Z_{2}} (z_{1},z_{2}) dz_{2} dz_{1} \nonumber \\
&=\int \limits_{0}^{t} \int \limits_{0}^{t-z_{1}} \frac{\left(4ab^{2}c \right)^{\frac{n-1}{2}}\Gamma \left (n-\frac{1}{2} \right) z_{1}^{\frac{n-1}{2}-1} z_{2}^{\frac{n-1}{2}-1} dz_{2} dz_{1}}{2^{n-\frac{1}{2}} \Gamma \left( \frac{1}{2}\right) \Gamma \left( \frac{n-1}{2}\right)^{2} \left((1-2z_{1} ) ( 1-2z_{2})\right)^{\frac{n+1}{2}}  \left(\frac{1}{2}+ \frac{abz_{1}}{(1-2z_{1})}+\frac{bcz_{2}}{(1-2z_{2})}\right)^{n-\frac{1}{2}}}  \nonumber \\
&=\int \limits_{0}^{t}  \frac{\left(4ab^{2}c \right)^{\frac{n-1}{2}}\Gamma \left (n-\frac{1}{2} \right) z_{1}^{\frac{n-1}{2}-1}}{2^{n-\frac{1}{2}} \Gamma \left( \frac{1}{2}\right) \Gamma \left( \frac{n-1}{2}\right)^{2} \left(1-2z_{1} \right)^{\frac{n+1}{2}}}\textbf{B} dz_{1}
\end{align}
where
\begin{align} \label{eq29}
\textbf{B}&=\int \limits_{0}^{t-z_{1}} \frac{z_{2}^{\frac{n-1}{2}-1} \left( 1-2z_{2}\right)^{\frac{n+1}{2}}dz_{2}}{\left(\frac{1}{2}+ \frac{abz_{1}}{(1-2z_{1})}+\frac{bcz_{2}}{(1-2z_{2})}\right)^{n-\frac{1}{2}}}
\end{align}

To compute the integral at Equation (\ref{eq29}), first we will simplify this $\\ \left(\frac{1}{2}+ \frac{abz_{1}}{(1-2z_{1})}+\frac{bcz_{2}}{(1-2z_{2})}\right)^{n-\frac{1}{2}}$  as $\left(1+\frac{2(bc-(1-2(1-ab-bc)z_{1}))}{1-2(1-ab)z_{1}} z_{2} \right) \left( \frac{1-2(1-ab)z_{1}}{2(1-2z_{1})}\right)$. Therefore the Equation (\ref{eq29}) becomes

\begin{align} \label{eq210}
\frac{\left(2(1-2z_{1})\right)^{n-\frac{1}{2}}}{\left( 1-2(1-ab)z_{1}\right)^{n-\frac{1}{2}}} \int \limits_{0}^{t-z_{1}} \frac{z_{2}^{\frac{n-1}{2}-1} \left( 1-2z_{2}\right)^{\frac{n+1}{2}}  dz_{2}} {\left( 1+\frac{2(bc-(1-2(1-ab-bc)z_{1}))}{1-2(1-ab)z_{1}} z_{2} \right)^{n-\frac{1}{2}}}
\end{align}

The integral $\int \limits_{0}^{t-z_{1}} \frac{z_{2}^{\frac{n-1}{2}-1} \left( 1-2z_{2}\right)^{\frac{n+1}{2}}  dz_{2}} {\left( 1+\frac{2(bc-(1-2(1-ab-bc)z_{1}))}{1-2(1-ab)z_{1}} z_{2} \right)^{n-\frac{1}{2}}} $ is written as
\begin{align} \label{eq211}
\int \limits_{0}^{t-z_{1}} z_{2}^{\frac{n-1}{2}-1} \left( 1-2z_{2}\right)^{\frac{n+1}{2}} \left( 1+\frac{2(bc-(1-2(1-ab-bc)z_{1}))}{1-2(1-ab)z_{1}} z_{2} \right)^{-\left(n-\frac{1}{2}\right)}dz_{2} 
\end{align}
By considering the binomial expansion then Equation (\ref{eq211}) can be represented as 
\begin{align} \label{eq212}
&=\sum \limits_{k=0}^{\infty} \sum \limits_{l=0}^{\infty} \Bigg[ \frac{2^{k+l} (-1)^{k+l} \dbinom{\frac{n+1}{2}}{k}\dbinom{n-\frac{3}{2} +l}{l}(bc-1+2(1-ab-bc)z_{1})^{l}}{\left( 1-2(1-ab)z_{1}\right)^{l}}  \times \nonumber \\
& \int \limits_{0}^{t-z_{1}} z_{2}^{\frac{n-1}{2}+k+l}dz_{2} \Bigg] \nonumber \\
&=\sum \limits_{k=0}^{\infty}\sum_{l=0}^{\infty} \Bigg[ \frac{2^{k+l} (-1)^{k+l} \dbinom{\frac{n+1}{2}}{k}\dbinom{n-\frac{3}{2} +l}{l}(bc-1+2(1-ab-bc)z_{1})^{l}}{\left( 1-2(1-ab)z_{1}\right)^{l}}  \times \nonumber \\
&\frac{\left(t-z_{1}\right)^{\frac{n+1}{2}+k+l}} {\frac{n+1}{2}+k+l} \Bigg]
\end{align}

Therefore,
\begin{align} \label{eq213}
\textbf{B}&=\frac{\left(2(1-2z_{1})\right)^{n-\frac{1}{2}}}{\left( 1-2(1-ab)z_{1}\right)^{n-\frac{1}{2}}} \times \nonumber \\
& \Bigg[ \sum \limits_{k=0}^{\infty}\sum_{l=0}^{\infty} \Bigg[ \frac{2^{k+l} (-1)^{k+l} \dbinom{\frac{n+1}{2}}{k}\dbinom{n-\frac{3}{2} +l}{l}(bc-1+2(1-ab-bc)z_{1})^{l}}{\left( 1-2(1-ab)z_{1}\right)^{l}}\times \nonumber \\
& \frac{\left(t-z_{1}\right)^{\frac{n+1}{2}+k+l}} {\frac{n+1}{2}+k+l} \Bigg] \Bigg]
\end{align}

and \\
\begin{align} \label{eq214}
F_{T} (t)&=\frac{\left(4ab^{2}c \right)^{\frac{n-1}{2}}\Gamma \left (n-\frac{1}{2} \right)}{\Gamma \left( \frac{1}{2}\right) \Gamma \left( \frac{n-1}{2}\right)^{2}}\times \nonumber \\
&\Bigg[\sum \limits_{k=0}^{\infty}\sum_{l=0}^{\infty}  \frac{(-2)^{k} \dbinom{\frac{n+1}{2}}{k}\dbinom{n-\frac{3}{2} +l}{l}(-2(bc-1))^{l}t^{\frac{n+1}{2}+k+l} }  {\frac{n+1}{2}+k+l}\textbf{G}\Bigg]
\end{align}

where
\begin{align} \label{eq215}
\textbf{G}&= \int \limits_{0}^{t}  \Bigg[z_{1}^{\frac{n-1}{2}-1} \left( 1-\frac{z_{1}}{t}\right)^{\frac{n+1}{2}+k+l} (1-2z_{1})^{\frac{n}{2}-1} \left( 1-2(1-ab)z_{1}\right)^{-(n+l-\frac{1}{2})}\times \nonumber \\
& \left(1+\frac{2(1-ab-bc)z_{1}}{bc-1}\right)^{l} dz_{1} \Bigg]
\end{align}

Again, by considering the binomial expansion then Equation (\ref{eq215}) can be written as follows
\begin{align}  \label{eq216}
\textbf{G}&=\sum \limits_{m=0}^{\infty} \sum_{p=0}^{\infty} \sum_{q=0}^{\infty} \Bigg[ \dbinom{\frac{n}{2}}{m} \dbinom{l}{p} \dbinom{n+l+q-\frac{3}{2}}{q} (-2)^{m} \left(\frac{2(1-ab-bc)}{bc-1} \right)^{p} (-2(1-ab))^{q} \times \nonumber \\
&\int \limits_{0}^{t}  z_{1}^{\frac{n-1}{2+m+p+q}-1} \left( 1-\frac{z_{1}}{t}\right)^{\frac{n-1}{2}+k+l}dz_{1} \Bigg] \nonumber \\
&=\sum \limits_{m=0}^{\infty} \sum_{p=0}^{\infty} \sum_{q=0}^{\infty} \Bigg[ \dbinom{\frac{n}{2}}{m} \dbinom{l}{p} \dbinom{n+l+q-\frac{3}{2}}{q} (-2)^{m} \left(\frac{2(1-ab-bc)}{bc-1} \right)^{p} (-2(1-ab))^{q} \times \nonumber \\
&\frac {B\left ( \frac{n-1}{2}+m+p+q,\frac{n+1}{2}+k+l \right)} {t^{\frac{n-1}{2}+m+p+q} } \Bigg]
\end{align}

Therefore $F_{T}(t)$ is 
\begin{align} \label{eq217}
F_{T} (t)&=\frac{\left(4ab^{2}c \right)^{\frac{n-1}{2}}\Gamma \left (n-\frac{1}{2} \right)}{\Gamma \left( \frac{1}{2}\right) \Gamma \left( \frac{n-1}{2}\right)^{2}} \qquad \times \nonumber \\
&\Bigg[ \sum \limits_{k=0}^{\infty}\sum_{l=0}^{\infty} \sum_{m=0}^{\infty} \sum_{p=0}^{\infty} \sum_{q=0}^{\infty} \Bigg[ \frac{2^{k} \dbinom{\frac{n-1}{2}+k}{k}\dbinom{n-\frac{3}{2} +l}{l}(-2(bc-1))^{l}t^{\frac{n-1}{2}+k+l} }  {\frac{n-1}{2}+k+l} \qquad \times \nonumber \\
&\frac{\dbinom{\frac{n}{2}}{m} \dbinom{l}{p} \dbinom{n+l+q-\frac{3}{2}}{q} (-2)^{m} \left(\frac{2(1-ab-bc)}{bc-1} \right)^{p} (-2(1-ab))^{q} } {t^{\frac{n-1}{2}+m+p+q}} \qquad \times \nonumber \\
&B \left(\frac{n-1}{2}+m+p+q,\frac{n+1}{2}+k+l \right) \Bigg] \Bigg]\nonumber \\
&=\frac{\left(4ab^{2}c \right)^{\frac{n-1}{2}}\Gamma \left (n-\frac{1}{2} \right)}{\Gamma \left( \frac{1}{2}\right) \Gamma \left( \frac{n-1}{2}\right)^{2}} \qquad \times \nonumber \\
&\Bigg[ \sum \limits_{k=0}^{\infty}\sum_{l=0}^{\infty} \sum_{m=0}^{\infty} \sum_{p=0}^{\infty} \sum_{q=0}^{\infty} \Bigg[ \frac{2^{k} \dbinom{\frac{n-1}{2}+k}{k}\dbinom{n-\frac{3}{2} +l}{l}(-2(bc-1))^{l}t^{k+l-m-p-q} }  {\frac{n-1}{2}+k+l} \qquad \times \nonumber \\
&\dbinom{\frac{n}{2}}{m} \dbinom{l}{p} \dbinom{n+l+q-\frac{3}{2}}{q} (-2)^{m} \left(\frac{2(1-ab-bc)}{bc-1} \right)^{p} (-2(1-ab))^{q} \qquad  \times \nonumber \\
&B \left(\frac{n-1}{2}+m+p+q,\frac{n+1}{2}+k+l \right) \Bigg] \Bigg]
\end{align}

Furthermore, from the Equation (\ref{eq217}) then the probability density function (pdf) of $T$ is
\begin{align} \label{eq218}
f_{T} (t)&=\frac{\left(4ab^{2}c \right)^{\frac{n-1}{2}}\Gamma \left (n-\frac{1}{2} \right)}{\Gamma \left( \frac{1}{2}\right) \Gamma \left( \frac{n-1}{2}\right)^{2}} \sum \limits_{k=0}^{\infty}\sum_{l=0}^{\infty} \sum_{m=0}^{\infty} \sum_{p=0}^{\infty} \sum_{q=0}^{\infty} \frac{2^{k} \dbinom{\frac{n-1}{2}+k}{k}\dbinom{n-\frac{3}{2} +l}{l}} {\frac{n-1}{2}+k+l} \quad \times \nonumber \\
&(k+l-m-p-q) t^{k+l-m-p-q} (-2(bc-1))^{l} \dbinom{\frac{n}{2}}{m} \dbinom{l}{p} \dbinom{n+l+q-\frac{3}{2}}{q} (-2)^{m} \times \nonumber \\
& \left(\frac{2(1-ab-bc)}{bc-1} \right)^{p} (-2(1-ab))^{q}  B \left(\frac{n-1}{2}+m+p+q,\frac{n+1}{2}+k+l \right) 
\end{align}

The pdf of $T$ also can be computed as follows:
\begin{align} \label{eq219}
f_{T} (t)&=\int \limits_{-\infty}^{t} f_{Z_{1},Z_{2}} (z_{1},t-z_{1}) dz_{1}\nonumber \\
&=\int \limits_{-\infty}^{t} f_{Z_{1},Z_{2}} (t-z_{2},z_{2}) dz_{2}
\end{align}

Now we already have the pdf and cdf of $T$ from where we can compute the $T$ statistics value for the hypothesis testing, that is the hypothesis null about the equality of mean of two-groups independent samples is rejected if $T \le t_{0}$ or $P \left(T \le t_{0} \right)=F_{T}(t_{0}) \le \alpha$. 

The computation of $F_{T}(t_{0})$ by using the Equation (\ref{eq218}) is involving the five summation and hence it is not quite simple. The computation gets simple in the case $\sigma_{x}^{2} = \sigma_{y}^{2} $, which will be described next.

\subsection{Special case of the proposed test} \label{sec23}

In the case of $\sigma_{x}^{2} = \sigma_{y}^{2} $, we estimate the $V_{x}$ and $V_{y}$ by the least square estimator of the pooled variance $S^{2}_{p} = \frac{V_{x}+V_{y}}{2}$.

If we use the least square estimate as the estimator of $V_{x}$ and $V_{y}$ therefore the Equations (\ref{eq23}) and (\ref{eq25}) becomes
\begin{subequations}
\begin{align}
T^{*} &= \frac{\frac{V_{x}+V_{y}}{2}}{\Big[\frac{V_{x}+V_{y}}{2} + \frac{n\left(\overline{Y}-\overline{X}\right)^{2}}{n-1}\Big]} \label{eq220a}\\
 &= \frac{U^{*}}{U^{*}+4V^{*}} \label{eq220b}
\end{align}
\end{subequations}
where $U^{*}=\frac{(n-1)\left(V_{x}+V_{y}\right)}{\sigma_{x}^{2}}$ and $V^{*}=\frac{n\left(\overline{Y}-\overline{X}\right)^{2}}{2\sigma_{x}^{2}}$.

The pdf of $T^{*}$ is derived from the ratio of linear combination of chi-square random variables \cite{Pro94}. First, let $Y=1+4 \frac{V^{*}}{U^{*}}$, where $V^{*}$ is distributed $\chi^{2}_{(1)}$ and $U^{*}$ is distributed $\chi^{2}_{2(n-1)}$. Second, the pdf of $T^{*}$ is computed by taking $T^{*} = \frac{1}{Y}$.

In the folowing computation, the chi-square distribution will be represented as the Gamma distribution. Therefore, we have that $V^{*}$ is Gamma distributed with parameters $\alpha_{1} =\frac{1}{2}$ and $\beta_{1}=2$.
$U^{*}$ is Gamma distributed with parameters $\alpha_{2} =(n-1)$ and $\beta_{2}=2$.
$U^{*}$ and $V^{*}$ are independents.

Suppose $G=\frac{V^{*}}{U^{*}}$ and if we take $U^{*}=H$, then we get $V^{*}= GH$. Further we got the Jacobian of this transformation variable random is $h$. Because $V^{*}$ and $U^{*}$ are independents, then the joint probability function of $G$ and $H$ is 
\begin{flalign}\label{eq221}
f_{G,H}(g,h)=f_{V^{*},U^{*}}(gh,h).h 
\end{flalign}
where
\begin{align} 
f_{V^{*},U^{*}}(gh,h)&=f_{V^{*}}(gh).f_{U^{*}}(h), \nonumber \\
f_{V^{*}}(gh)&=\frac{(gh)^{\alpha_{1}-1}e^{-\frac{gh}{\beta_{1}}}}{\beta_{1}^{\alpha_{1}} \Gamma({\alpha_{1}})}, \textrm{and  } \nonumber  \\ 
f_{U^{*}}(h)&=\frac{(h)^{\alpha_{2}-1}e^{-\frac{h}{\beta_{2}}}}{\beta_{2}^{\alpha_{2}} \Gamma({\alpha_{2}})} \nonumber
\end{align}
Therefore 
\begin{align} \label{eq222}
f_{G,H}(g,h)
&=\frac{(gh)^{\alpha_{1}-1}e^{-\frac{gh}{\beta_{1}}}}{\beta_{1}^{\alpha_{1}} \Gamma({\alpha_{1}})}.\frac{(h)^{\alpha_{2}-1}e^{-\frac{h}{\beta_{2}}}}{\beta_{2}^{\alpha_{2}} \Gamma({\alpha_{2}})} h \nonumber \\
&=\frac{g^{\alpha_{1}-1} h^{\alpha_{1}+\alpha_{2}-1} e^{(-\frac{(1+g)}{\beta}h)}}{\beta^{\alpha_{1}+\alpha_{2}} \Gamma({\alpha_{1}}) \Gamma(\alpha_{2})}, \quad \beta_{1}=\beta_{2}=\beta
\end{align}
We want to determine the pdf of $g$ then
\begin{align} \label{eq223}
f_{G}(g)
&= \int \limits_{0}^{\infty} f_{G,H}(g,h) dh \nonumber \\
&= \frac{g^{\alpha_{1}-1}}{B\left(\alpha_{1},\alpha_{2}\right) (1+g)^{\alpha_{1}+\alpha_{2}}} \nonumber \\
&= \frac{g^{\frac{1}{2}-1}}{B\left(\frac{1}{2},n-1\right) (1+g)^{n-\frac{1}{2}}} 
\end{align}

$G$ is distributed beta of second kind.

The next step is determining the distribution of \textbf{$Y$}. We define \textbf{$Y$}$=1+4G$ and by using the transformation random variable, where $G=\frac{Y-1}{4}$,  the pdf of \textbf{$Y$} is computed as follow
\begin{align} \label{eq224}
f_{Y}
&=\frac{\left(\frac{y-1}{4}\right)^{\frac{1}{2}-1}}{B\left(\frac{1}{2},n-1\right) \left(1+\left(\frac{y-1}{4}\right )\right)^{n-\frac{1}{2}}}\frac{1}{4} \nonumber \\
&=\frac{4^{n-1} (y-1)^{\frac{1}{2}-1}}{B\left(\frac{1}{2},n-1\right)(y+3)^{n-\frac{1}{2}}}, \quad 1 \leq y \leq \infty
\end{align}

The pdf of $T^{*}=\frac{1}{Y}$ is obtained
\begin{align} \label{eq225}
f_{T^{*}}(t^{*})
&=\frac{4^{n-1} (\frac{1}{t^{*}}-1)^{\frac{1}{2}-1}}{B\left(\frac{1}{2},n-1\right)(3+\frac{1}{t^{*}})^{n-\frac{1}{2}}}\frac{1}{t^{*2}} \nonumber \\
&=\frac{4^{n-1} t^{n-2} (1-t^{*})^{\frac{1}{2}-1}}{B\left(\frac{1}{2},n-1\right)(1+3t^{*})^{n-\frac{1}{2}}}, \quad 0 \leq t^{*} \leq 1
\end{align}

Furthermore the cdf of $T^{*}$ analtically is computed as 
\begin{align} \label{eq226}
F_{T^{*}}(t^{*}_{0})&=\int \limits_{0}^{t^{*}_{0}} f_{T^{*}}(t^{*})dt^{*} \nonumber \\
&=\frac{3^{n-1}}{B\left(\frac{1}{2},n-1\right)} \int \limits_{0}^{t^{*}_{0}}  \frac{t^{*(n-2)} (1-t)^{\frac{1}{2}-1}}{(1+3t^{*})^{n-\frac{1}{2}}} dt^{*}  \nonumber \\
&=\frac{4^{n-1}}{B\left(\frac{1}{2},n-1\right)}  \sum \limits_{k=0}^{\infty} \left(-3\right)^k \binom{n-\frac{1}{2}+k-1}{k} \int \limits_{0}^{t^{*}_{0}}  t^{*(n-1+k-1)} (1-t^{*})^{\frac{1}{2}-1}dt^{*} \nonumber \\
&=\frac{4^{n-1}}{B\left(\frac{1}{2},n-1\right)} \Bigg[\sum \limits_{k=0}^{\infty} \left(-3\right)^k \binom{n+k-\frac{3}{2}}{k} B\left(n-1+k,\frac{1}{2}\right) \times \nonumber \\
& \textrm{pbeta}\left(t^{*}_{0},n-1+k,\frac{1}{2}\right) \Bigg]
\end{align}

We reject the null hypothesis of the equality of mean of two independent samples if $T^{*} < t^{*}_{0,\alpha}$ or $P \left(t^{*} < T^{*}_{0} \right)=F_{T^{*}}(T^{*}_{0}) =\text{p-value} \le P \left(t^{*} < t^{*}_{0,\alpha} \right)=F_{T^{*}}(t^{*}_{0,\alpha}) = \alpha$.

Observing that $(2(n-1)) \left(\frac{V^{*}}{U^{*}}\right)$ is the square of a random variable having $t_{2(n-1)}$ distribution, a simple calculation shows that the same holds for the random variable
\begin{eqnarray} \label{eq227}
J&=&\sqrt{(n-1) \left( \frac{1}{T^{*}} -1\right)} 
\end{eqnarray}
This statistic $J$ can also be used to test the hypothesis $\mu_{x}=\mu_{y}$ and the critical values can be computed from the $t$ table. It also follows that the $J$ has a limiting normal distribution as $n \rightarrow \infty$ 
\section{Simulation study}\label{sec3}
In this section we will describe the results from the two simulation studies of the $t$ and the proposed tests. First simulation is done to measure the power of the proposed test is conducted at $N=500$ times and $\alpha=0.01$. The results are presented at Section \ref{sec31}. In this simulation, we consider various possibilities regarding the sample sizes and variances. The results of the simulation are divided into groups according to
\begin{enumerate}
\item {variance: low ($S=0.2$), medium ($S=1.2$) and high ($ S=7$), }
\item{sample size: low ($5$ and $25$), medium ($100$) and high ($500$) }
\end{enumerate}
Section \ref{sec32} describes the second simulation results on measuring the rejection's rate under the null hypothesis of the $t$ and the proposed tests, based on the $N=500$ times simulation and $\alpha=0.01$. In the simulation, we choose $\mu_{X}=\mu_{Y}=\mu=9.2$ and $\sigma_{X}=\sigma_{Y}=\sigma$ are three values of variance $1.25, 3.5$ and $10$ which represents the low, medium and high variance respectively. 

Section \ref{sec33} gives some examples on how the proposed test works on the data.

\subsection{Power of the test} \label{sec31}

The simulation study to compute the power of the proposed and the $t$ tests is conducted as follows:

\begin{enumerate}
\item{Choose the $\mu_{X}, \mu_{Y},\sigma_{X}=\sigma_{Y}=\sigma$ of the two-groups independent samples}
\item{The simulation under null hypothesis
\begin{enumerate}
\item{Generate $n$ random samples normally distributed with mean and standard deviation, $\mu_{X_{0}}$ and $\sigma$}
\item{Generate $n$ random samples normally distributed with mean and standard deviation, $\mu_{Y_{0}}$ and $\sigma$}
\item{Compute their mean and variance samples}
\item{Compute $t^{*}_{0}$ values, based on the equation (\ref{eq220a}).}
\end{enumerate}}
\item{The simulation under alternative hypothesis
\begin{enumerate}
\item{Generate $n$ random samples normally distributed with mean and standard deviation, $\mu_{X_{1}}$ and $\sigma$}
\item{Generate $n$ random samples normally distributed with mean and standard deviation, $\mu_{Y_{1}}$ and $\sigma$}
\item{Compute their mean and variance samples}
\item{Compute $t^{*}_{1}$ values, based on the equation (\ref{eq220a}).}
\end{enumerate}}
\item{Repeat steps (2) - (3), $M$ times}
\item{Compute $t^{*}_{0,\alpha}$, the $\alpha$ quantile of $t^{*}_{0}$.  Note that $t^{*}_{0,\alpha}$ also can be computed by using the $\alpha^{th}$ quantile of pdf of $T^{*}$ in the Equation (\ref{eq221}).}
\item{Compute the power test of the proposed method $=1- \frac{\textrm{sum} (t^{*}_{1} \ge t^{*}_{0,\alpha})}{M}$}
\item{Compute the power of $t$ test from samples $X_{1}$ and $Y_{1}$} 
\item{Compare the results from steps (6) and (7)}
\item{Do steps (1)-(8), for different value of mean and variance}
\end{enumerate}

\begin{figure}[!h]
\begin{center}$
\begin{array}{ccc}
\includegraphics[width=1.5in,height=1.75in]{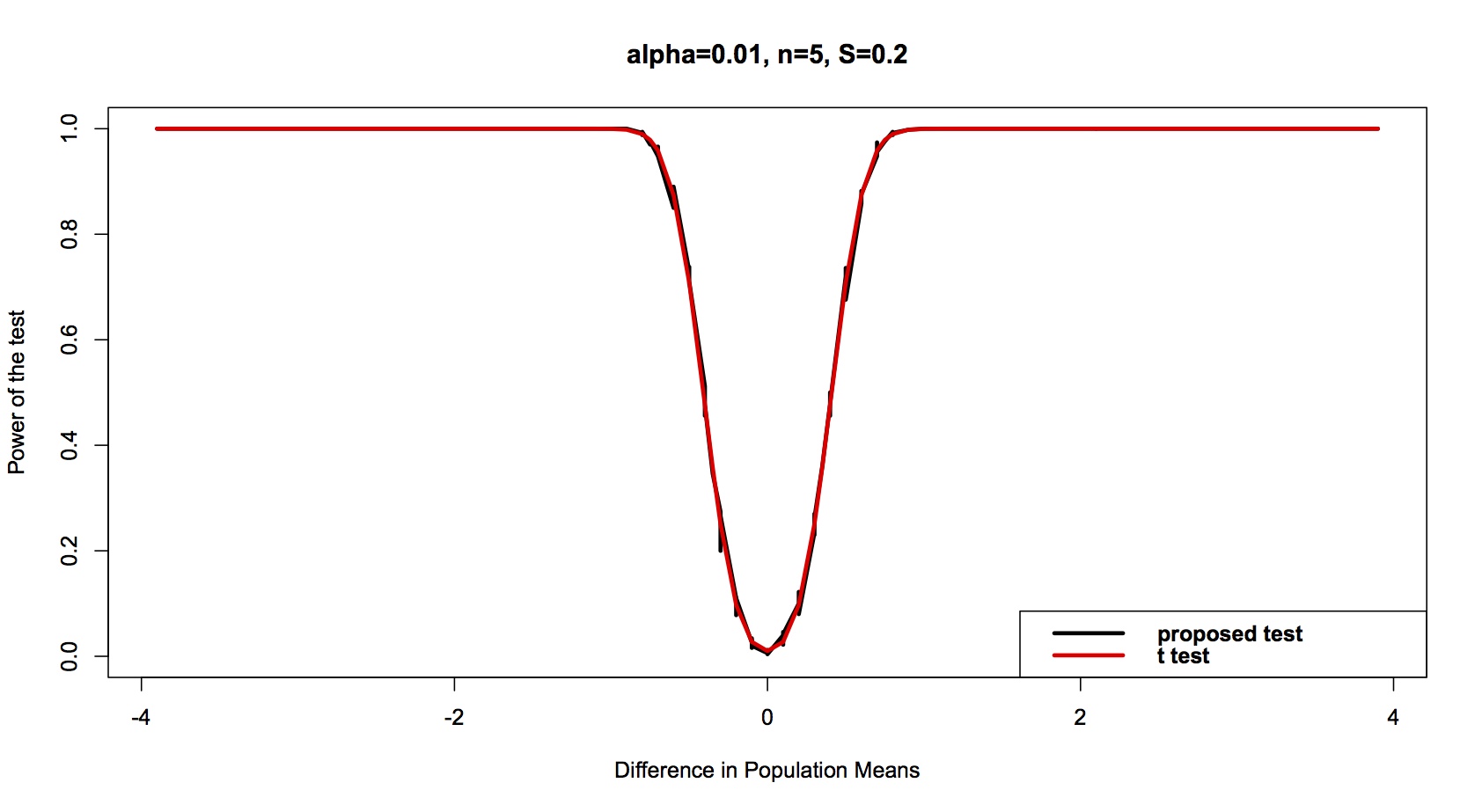}&
\includegraphics[width=1.5in,height=1.75in]{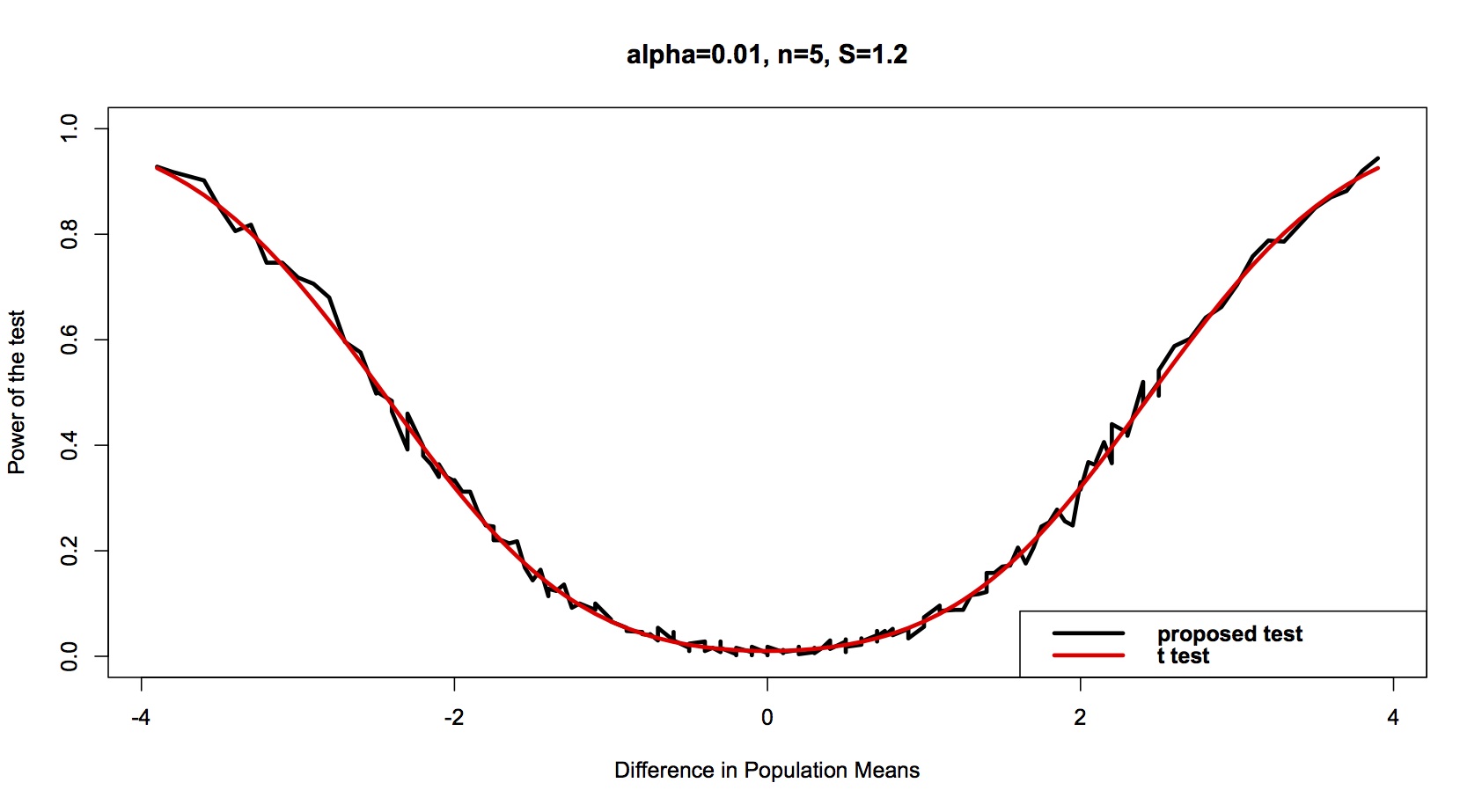}&
\includegraphics[width=1.5in,height=1.75in]{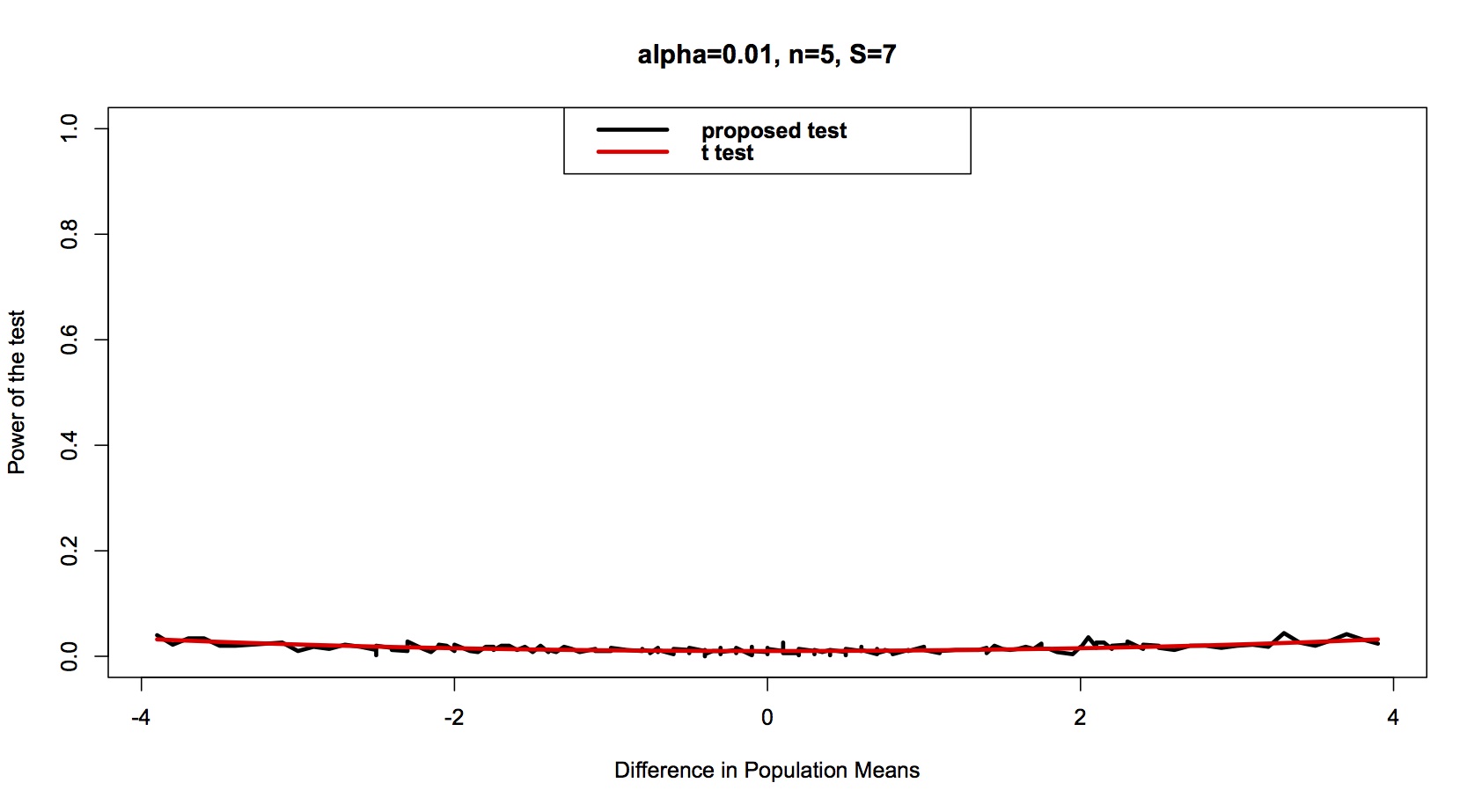} 
\end{array}$
\end{center}
\caption{Graphical power both the $t$ and proposed tests, n=5} \label{fig1}
\end{figure}

\begin{figure}[!h]
\begin{center}$
\begin{array}{ccc}
\includegraphics[width=1.5in,height=1.6in]{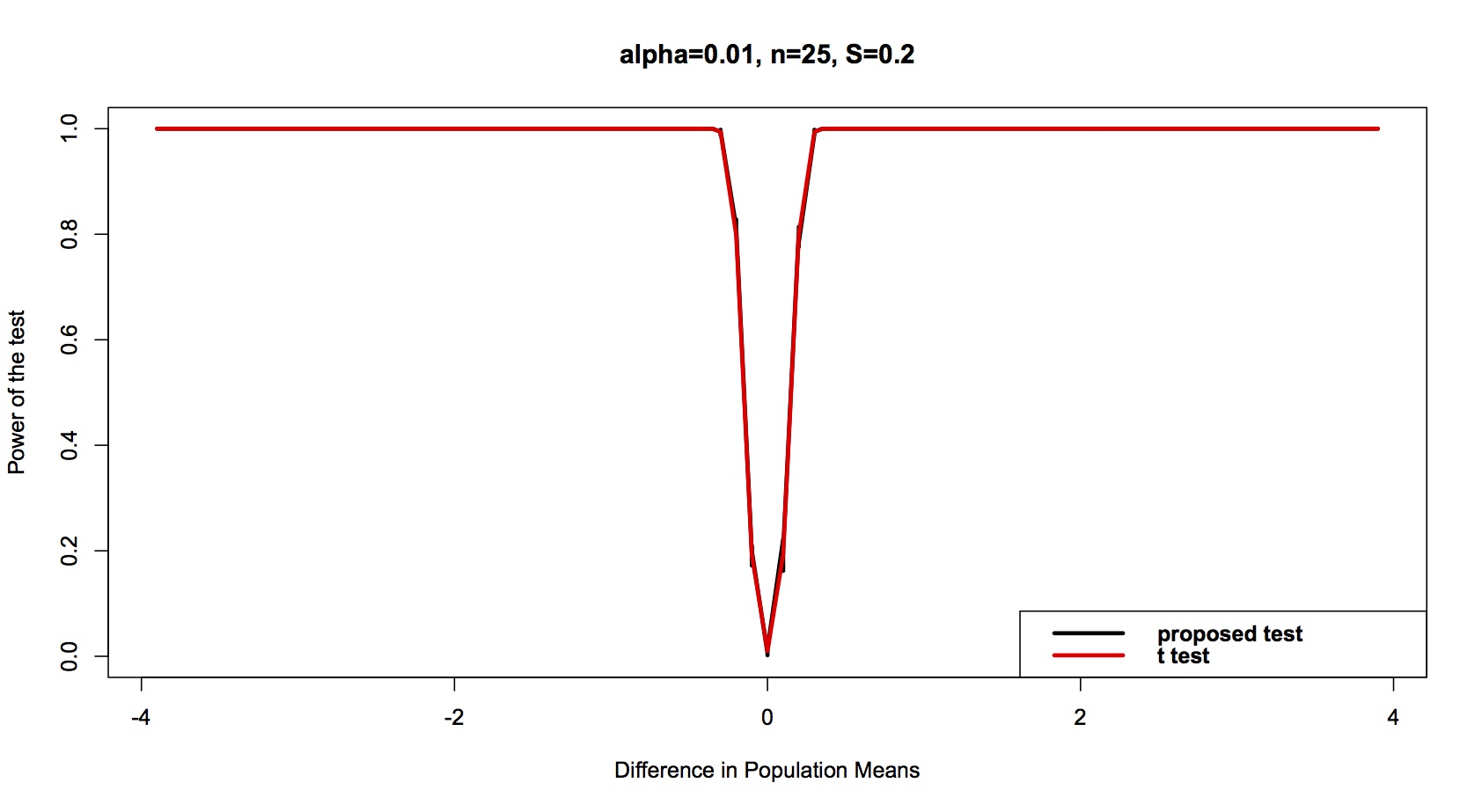}&
\includegraphics[width=1.5in,height=1.6in]{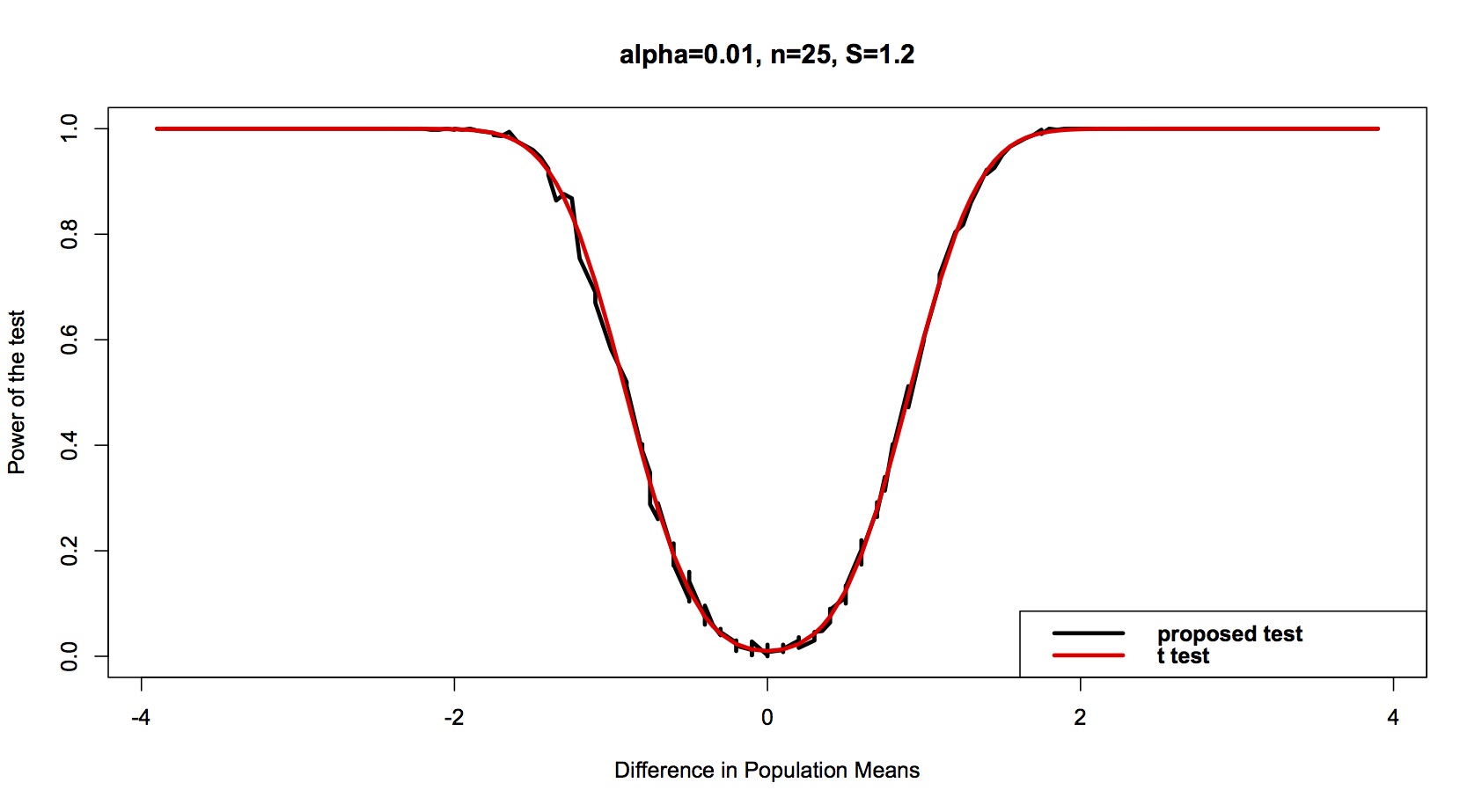}&
\includegraphics[width=1.5in,height=1.6in]{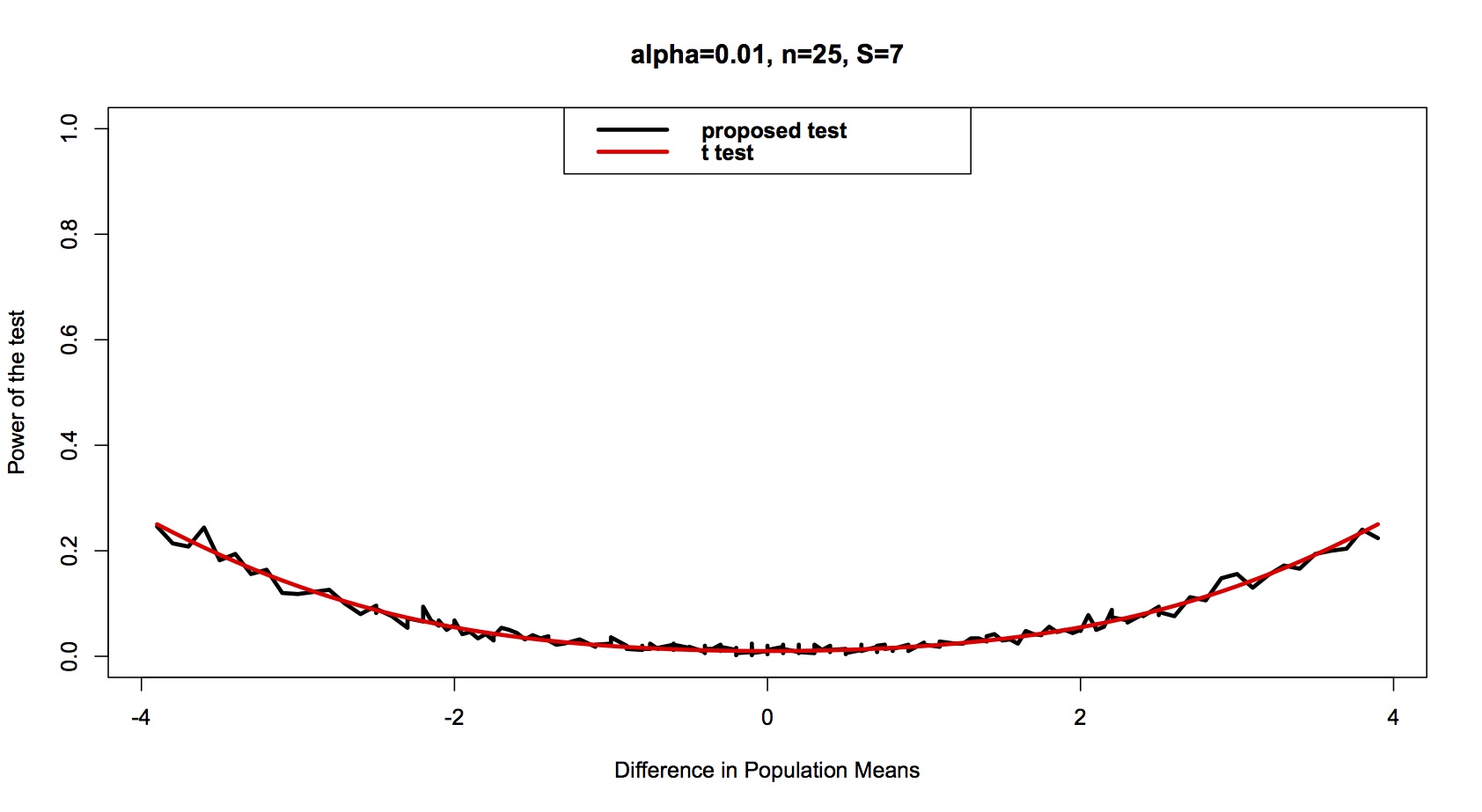} 
\end{array}$
\end{center}
\caption{Graphical power both the $t$ and proposed tests, n=25} \label{fig2}
\end{figure}

\begin{figure}[!h]
\begin{center}$
\begin{array}{ccc}
\includegraphics[width=1.5in,height=1.6in]{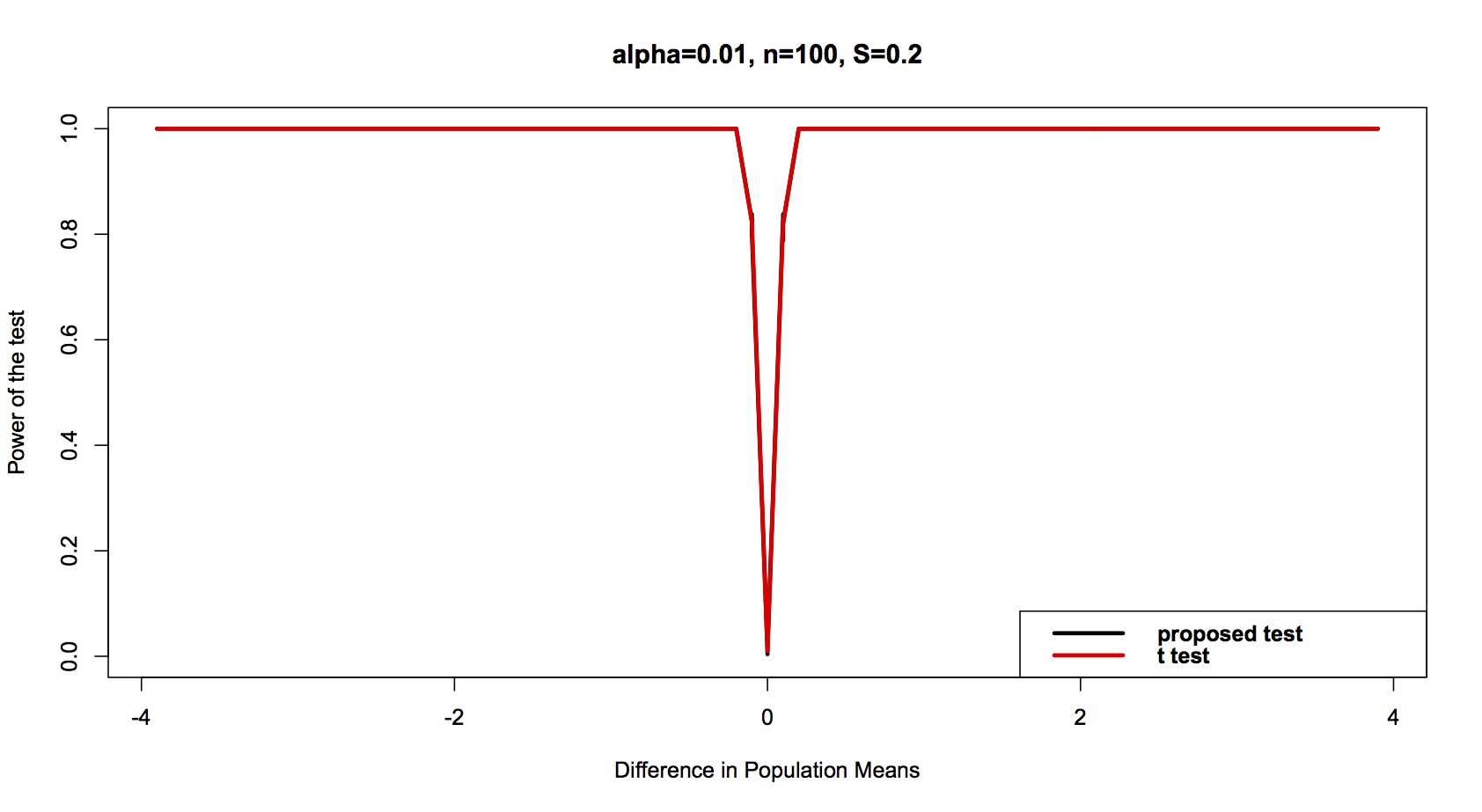}&
\includegraphics[width=1.5in,height=1.6in]{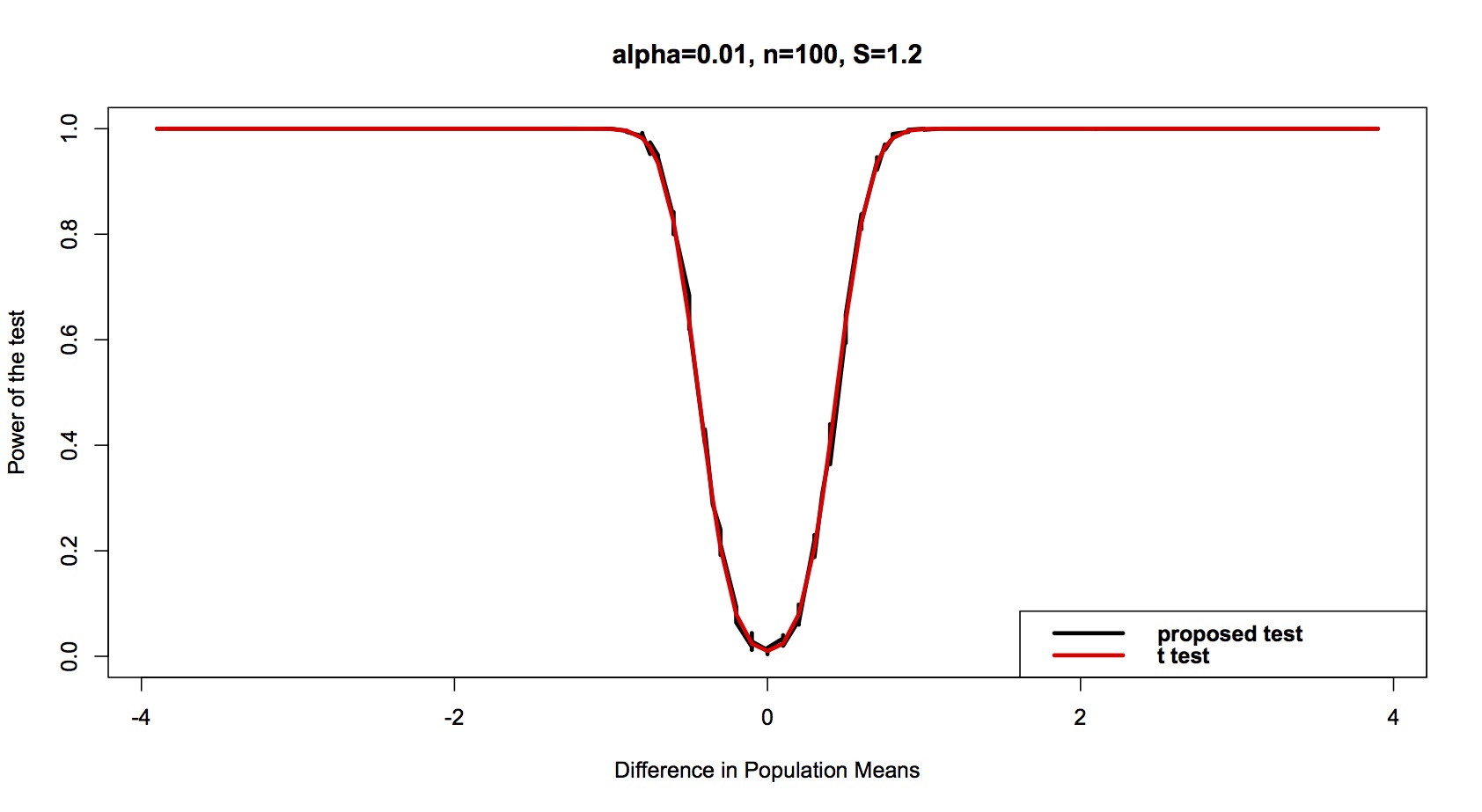}&
\includegraphics[width=1.5in,height=1.6in]{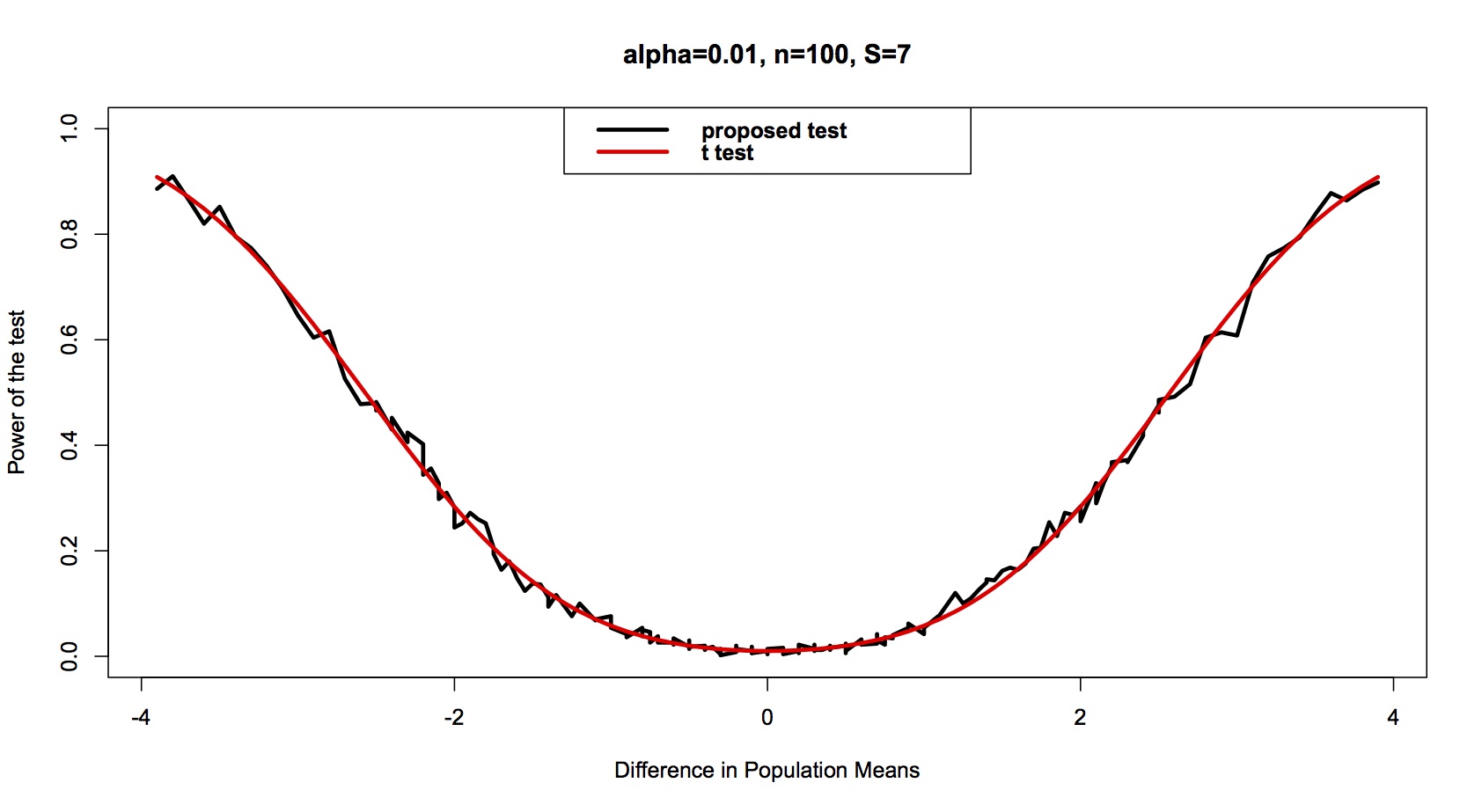} 
\end{array}$
\end{center}
\caption{Graphical power both the $t$ and proposed tests, n=100} \label{fig3}
\end{figure}

\begin{figure}[!h]
\begin{center}$
\begin{array}{ccc}
\includegraphics[width=1.5in,height=1.6in]{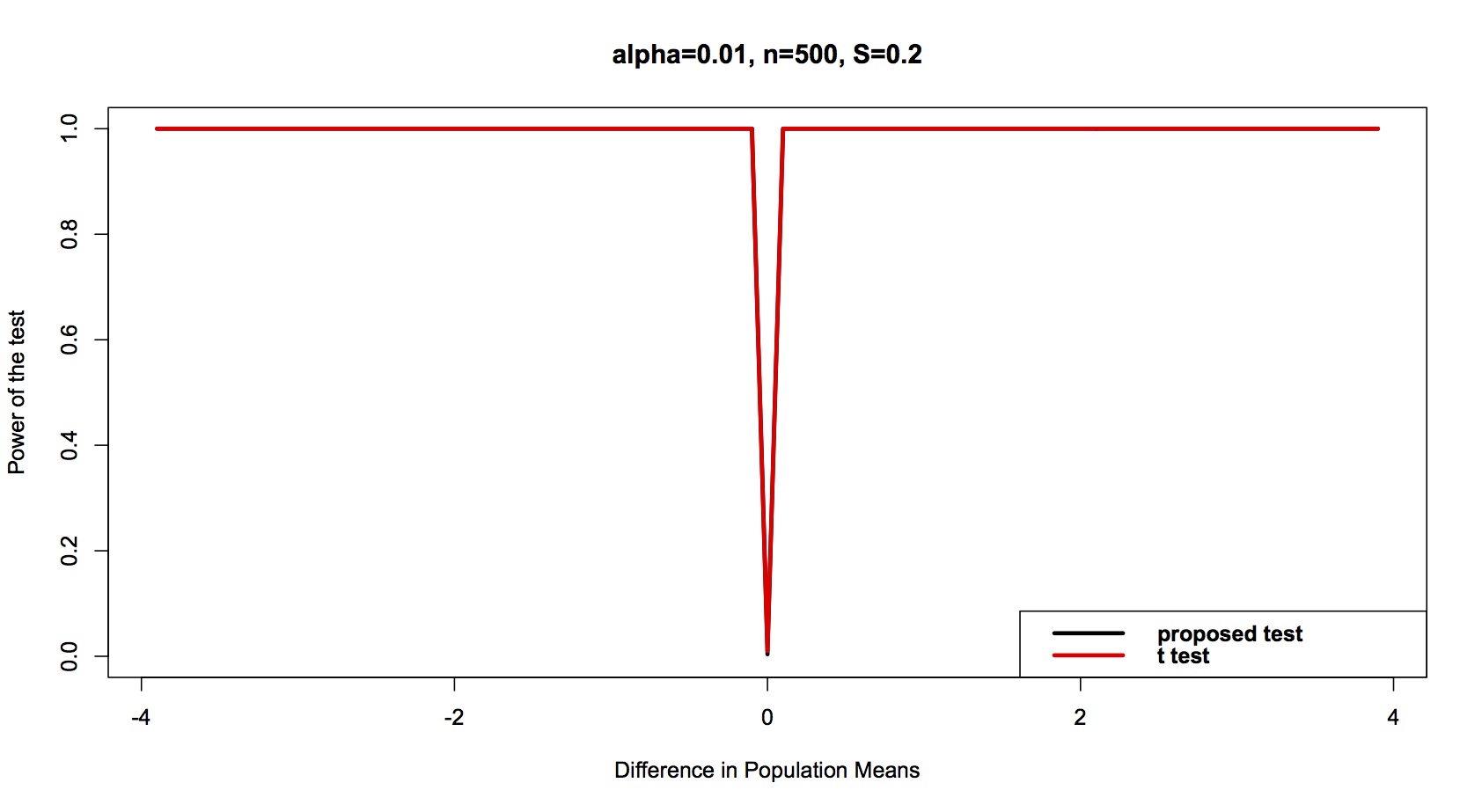}&
\includegraphics[width=1.5in,height=1.6in]{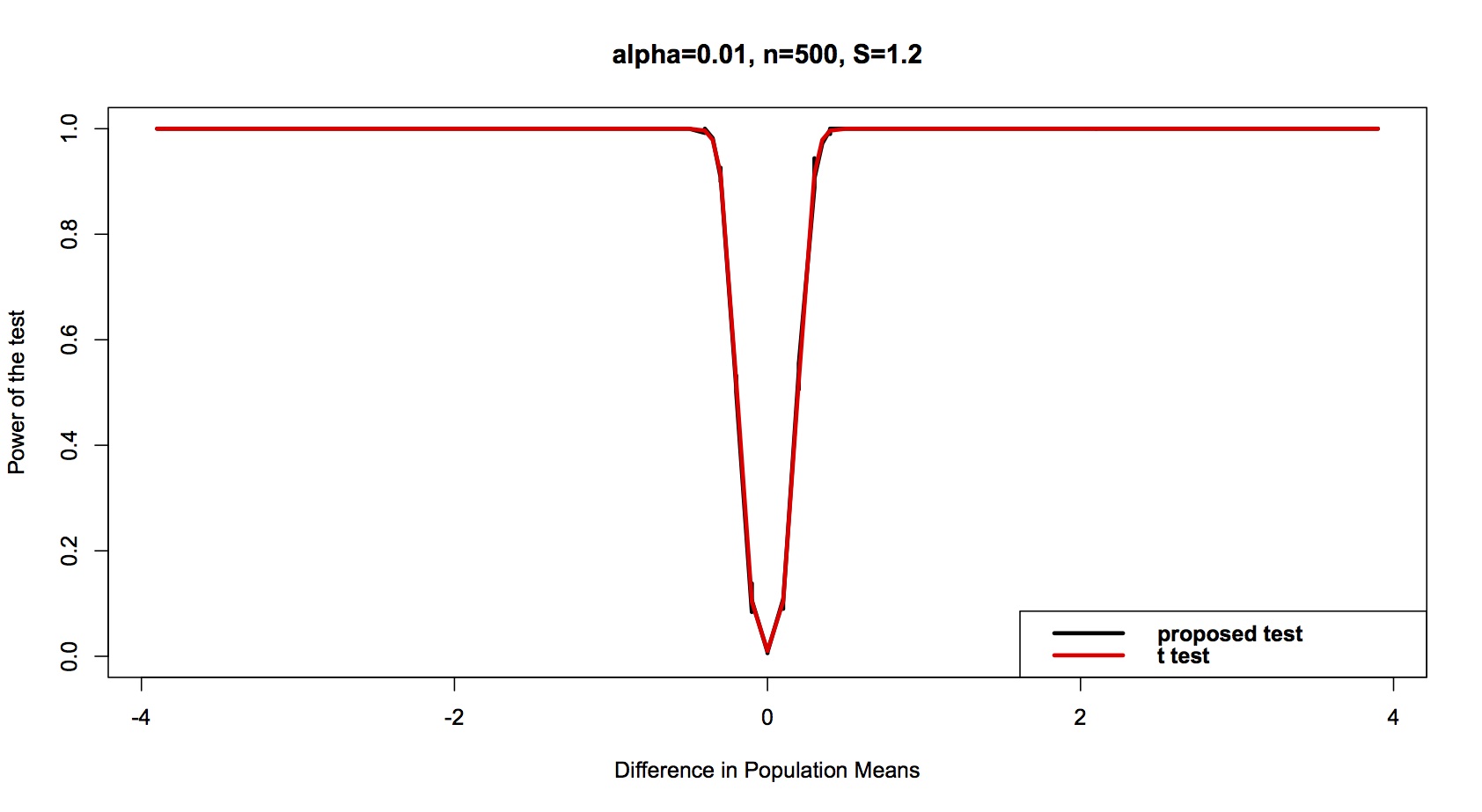}&
\includegraphics[width=1.5in,height=1.6in]{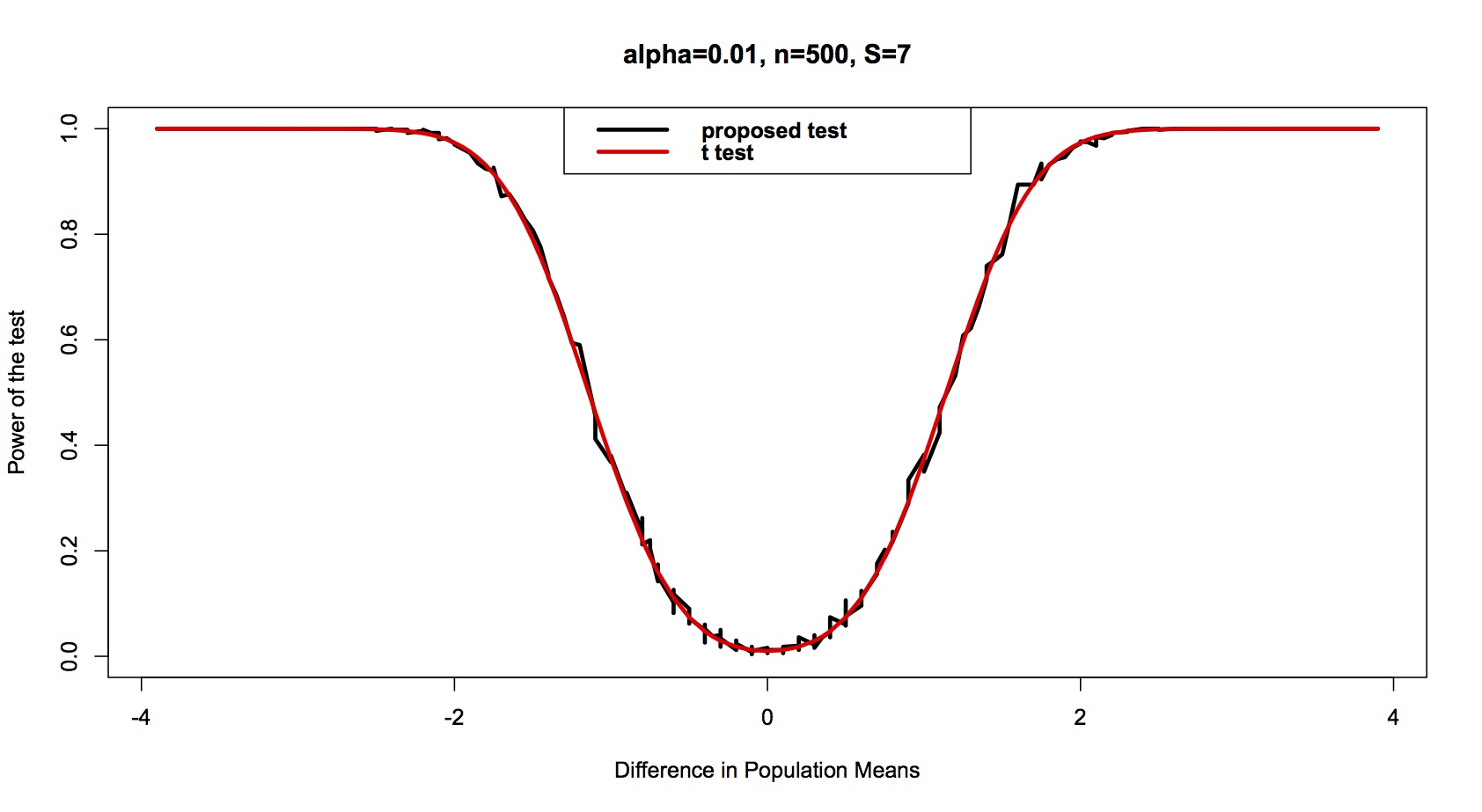} 
\end{array}$
\end{center}
\caption{Graphical power both the $t$ and proposed tests, n=500} \label{fig4}
\end{figure}

Figures \ref{fig1}, \ref{fig2}, \ref{fig3} and \ref{fig4} show the power of the special case of the cross variance and $t$ tests, based on the simulation study. They show that the proposed and the $t$ tests have an equal power. In the computation, the power of the proposed test is provided by implementing the the empirical approach, therefore the results is not as smooth as the $t$ test.

\newpage{}

\subsection{Error type I rate} \label{sec32}

The simulation study to compute the error type I rate is conducted as follows:

\begin{enumerate}
\item{Choose the $n, M_{1}, \alpha, \mu_{X}=\mu_{Y}=\mu$ and $\sigma_{X}=\sigma_{Y}=\sigma$ of the two-groups independent samples}
\item{Repeat $M_{1}$ times
	\begin{enumerate}
	\item{Compute p-value of $t^{*}_{o}$ test}
	\item{Compute p-value of $t$ test}
	\end{enumerate}}
\item{Compute the proportion of $t$ test p-value < $\alpha$ in $M_{1}$ results}
\item{Compute the proportion of the proposed test p-value < $\alpha$ in $M_{1}$ results}
\item{Compare the result of steps $3$ and $4$}
\end{enumerate}

In Table \ref{Tab1}, it is shown the results of the simulation where the error type I rate of the proposed and the $t$  tests are equal. 

\begin{table}[!h]
\begin{center}
\caption{Error type I rate under $500$ simulation for the proposed and $t$ tests} \label{Tab1}
\resizebox{8cm}{3cm}{
\begin{tabular}{clrr|rr}
\hline
\multirow{2}{*}{Sample size} 	& \multirow{2}{*}{Variance} 	&  \multicolumn{2}{c|}{proposed test} & \multicolumn{2}{|c}{$t$ test}  \\ \hhline{~~----}
	& & 0.05&0.01&0.05&0.01 \\ \hline
&low& 0.056 &0.012 &0.056 &0.012  \\ 
5&medium& 0.062 &0.016 &0.062 &0.016  \\
&high& 0.056 &0.020 &0.056 &0.020  \\ \hline
&low& 0.046 &0.010 &0.046 &0.010  \\ 
25&medium& 0.044 &0.012 &0.044 &0.012  \\ 
&high& 0.038 &0.010 &0.038 &0.010  \\ \hline
&low& 0.058 &0.006 &0.058 &0.006  \\ 
100&medium& 0.050 &0.010 &0.050 &0.010  \\ 
&high& 0.062 &0.012 &0.062 &0.012  \\  \hline
&low& 0.038 &0.004 &0.038 &0.004  \\ 
500&medium& 0.038 &0.002 &0.038 &0.002  \\ 
&high& 0.050 &0.010 &0.050 &0.010  \\ \hline
\end{tabular}
}
\end{center}
\end{table} 

\begin{figure}[!h]
        \centering
        \subfloat[n=5]{\includegraphics[width=2.5in,height=1.5in]{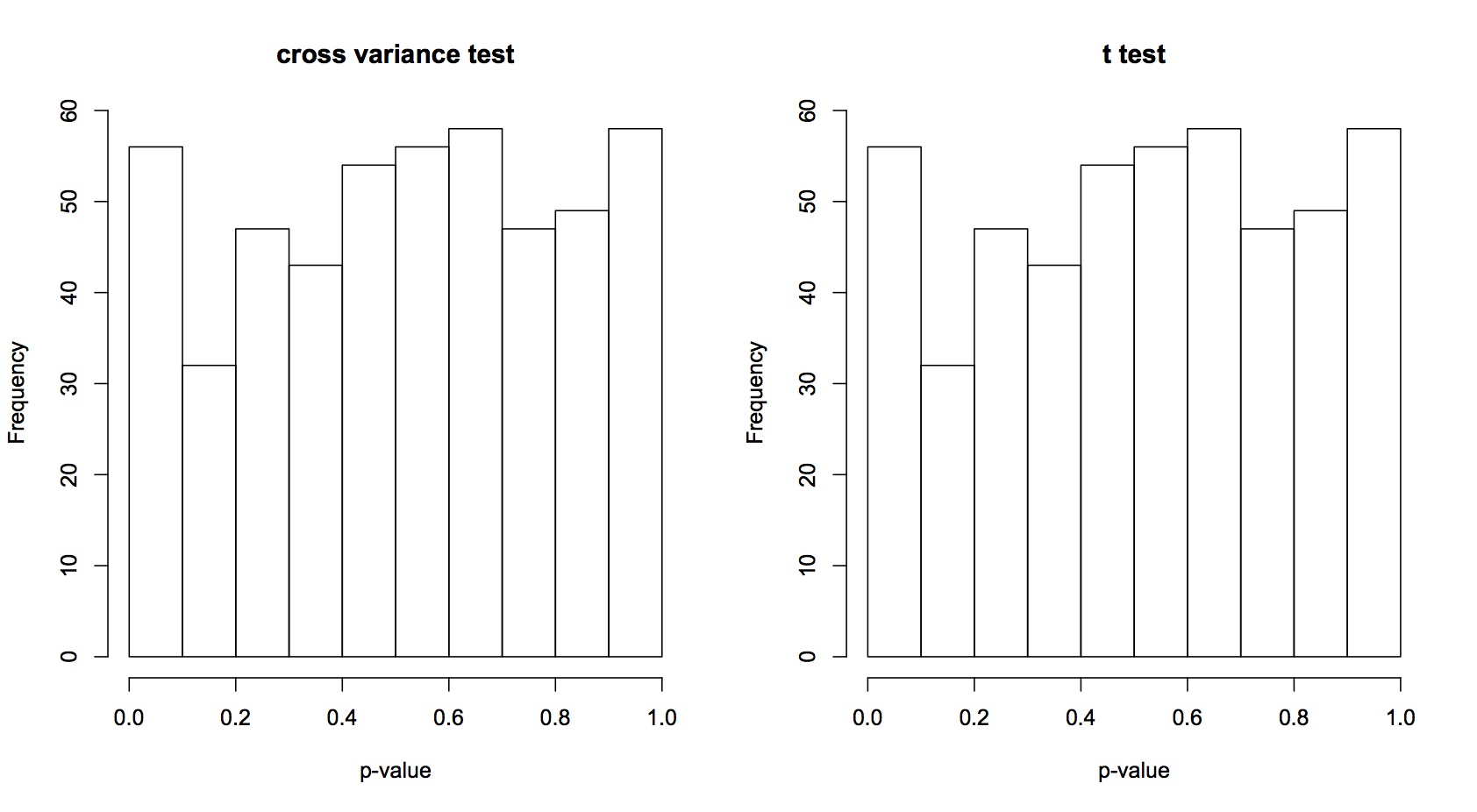}}
        \subfloat[n=500]{\includegraphics[width=2.5in,height=1.5in]{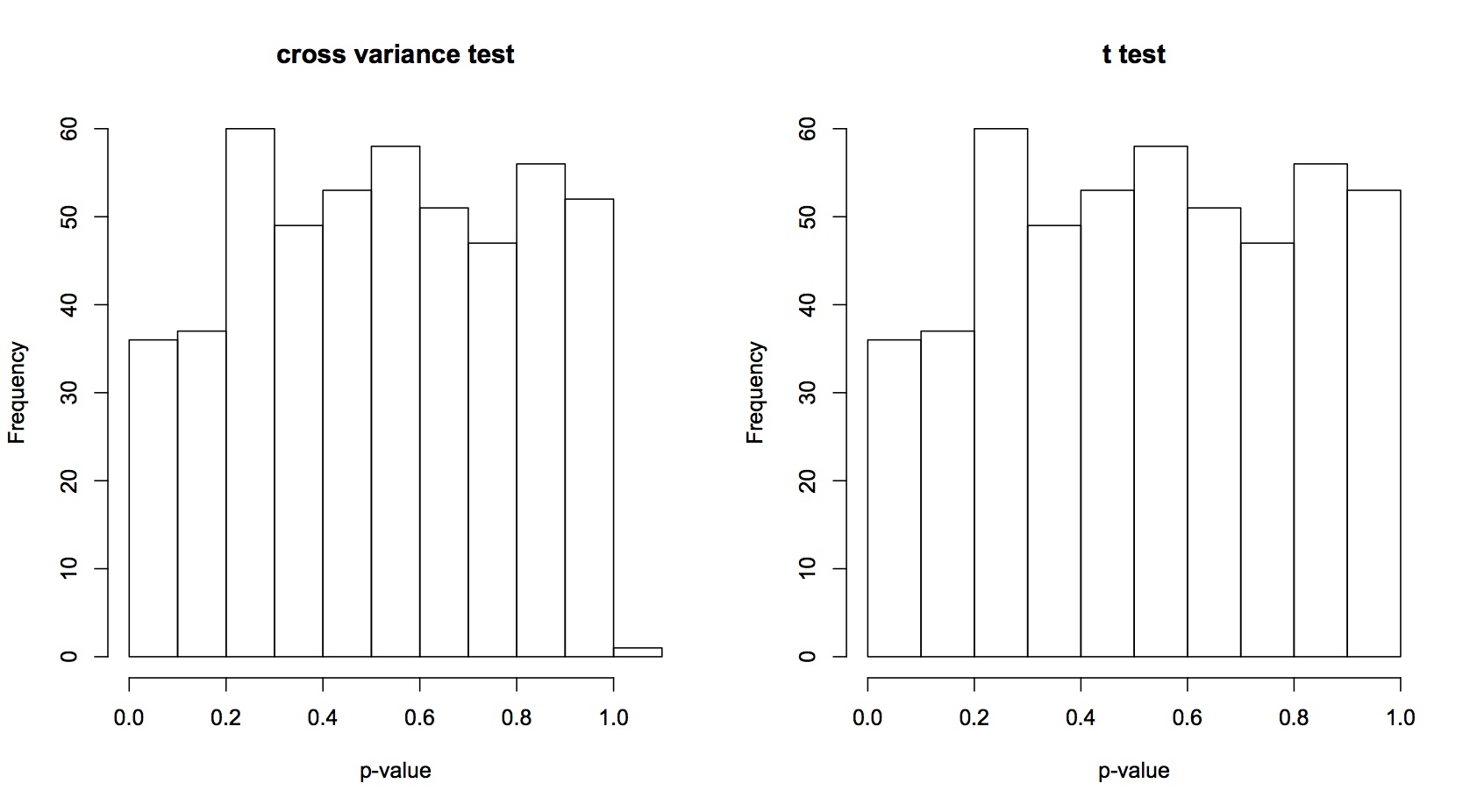}}
\caption{P-values distribution of the proposed and $t$ tests, small variance} \label{fig5}
 \end{figure}

\begin{figure}[!h]
        \centering
        \subfloat[n=5]{\includegraphics[width=2.5in,height=1.5in]{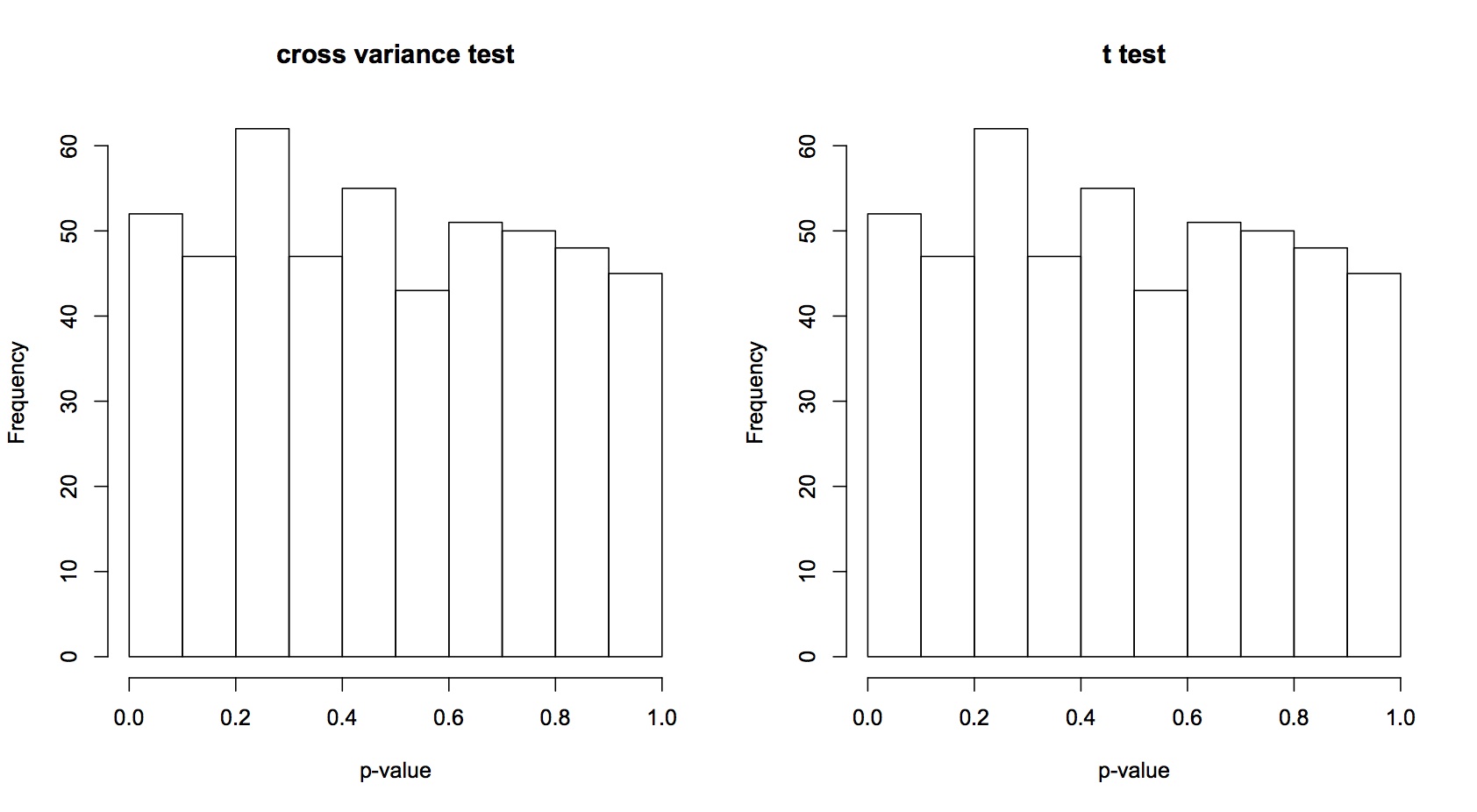}}
        \subfloat[n=500]{\includegraphics[width=2.5in,height=1.5in]{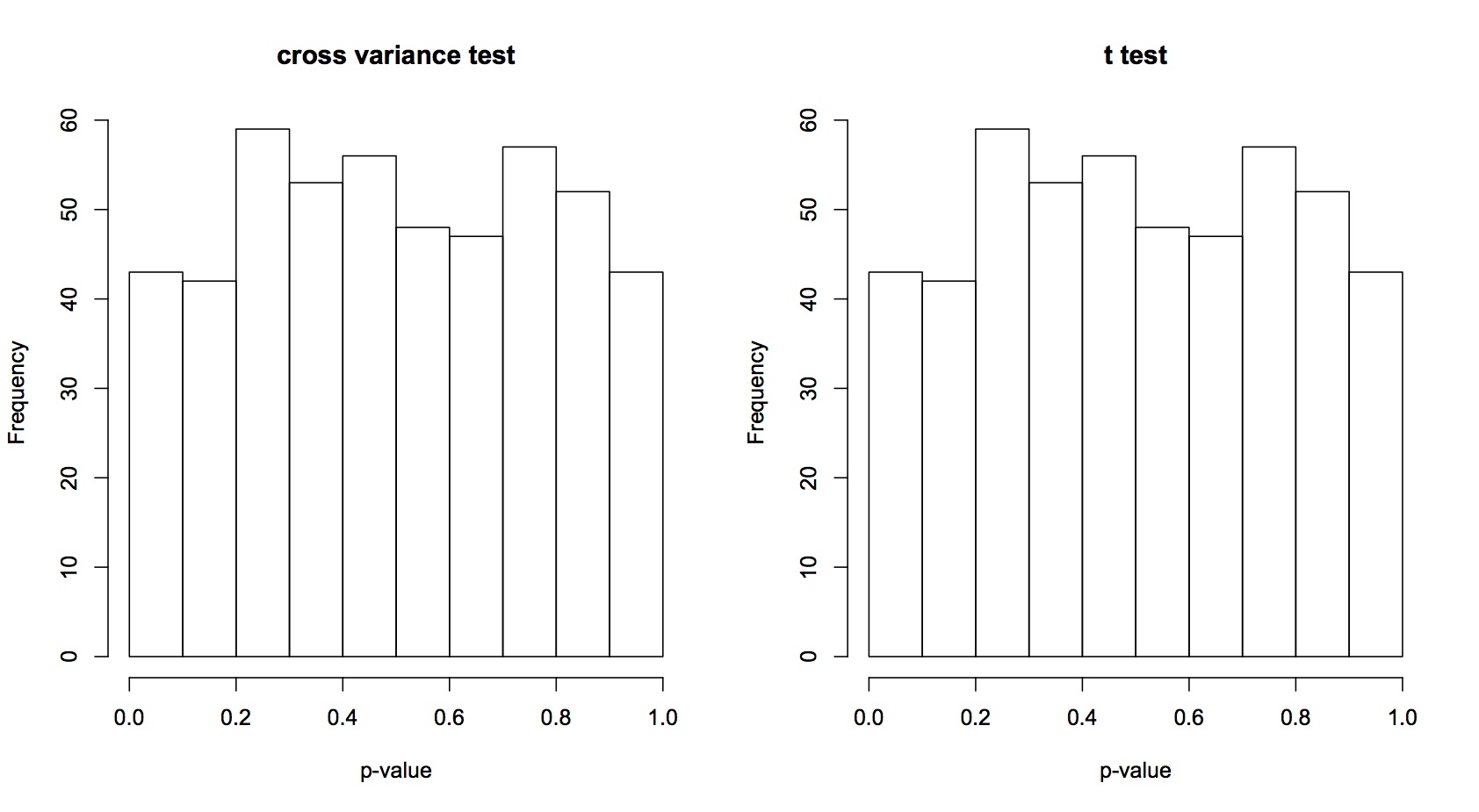}}
\caption{P-values distribution of the proposed and $t$ tests, medium variance} \label{fig6}
 \end{figure}

\begin{figure}[!h]
       \centering
        \subfloat[n=5]{\includegraphics[width=2.5in,height=1.5in]{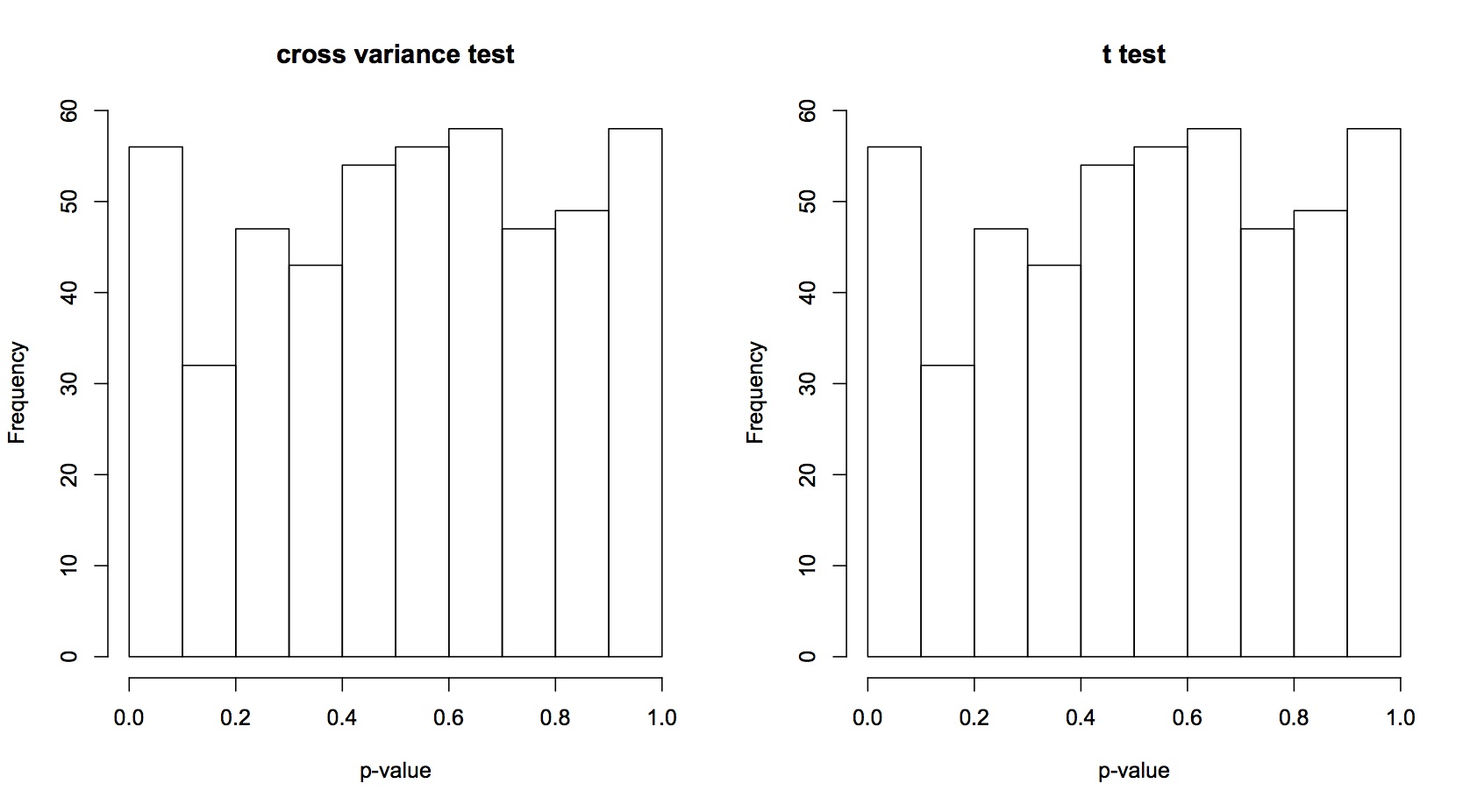}}
        \subfloat[n=500]{ \includegraphics[width=2.5in,height=1.5in]{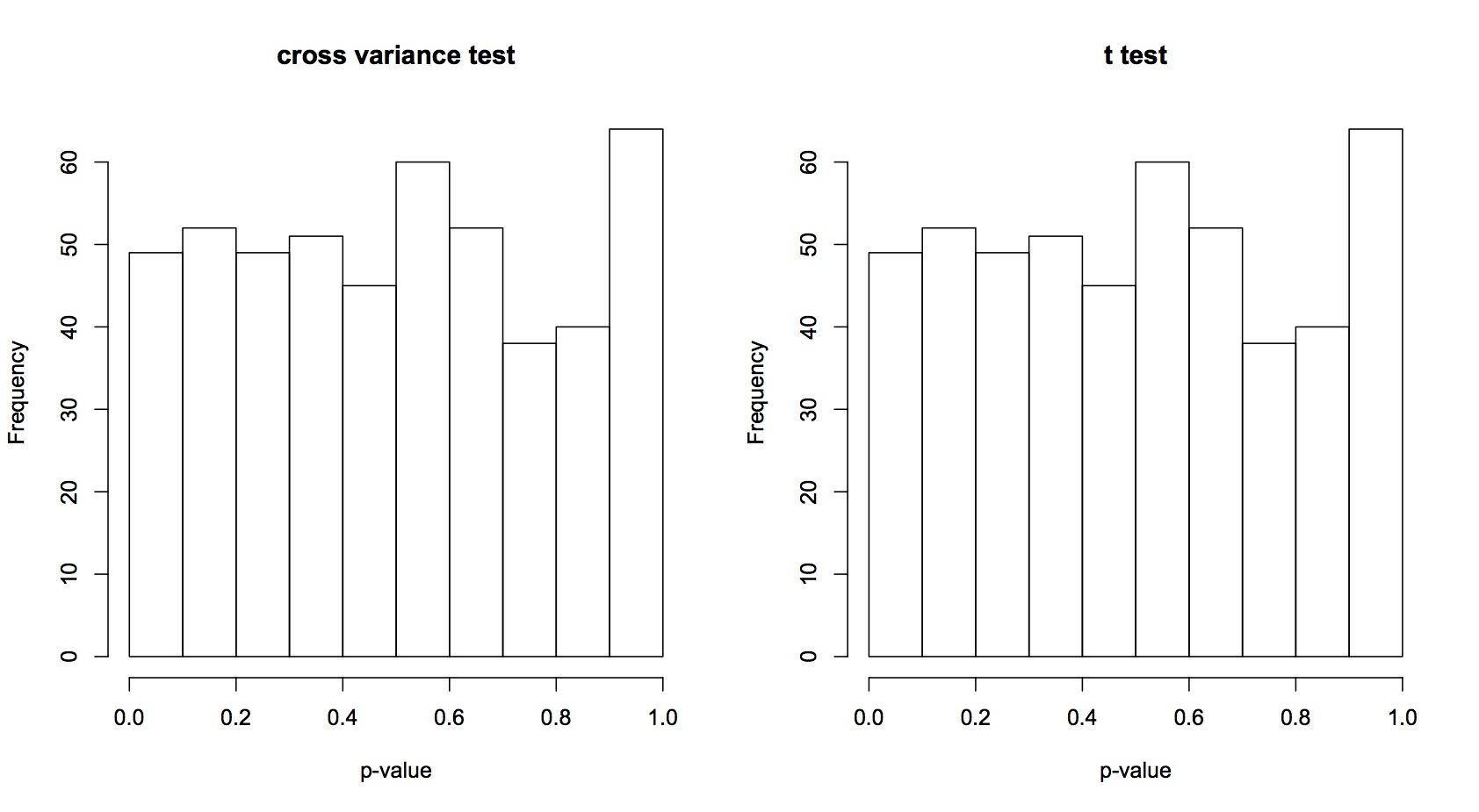}}
\caption{P-values distribution of the proposed and $t$ tests, high variance} \label{fig7}
 \end{figure}

The distribution of p-values from the proposed and $t$ tests are can be seen in the Figures \ref{fig5}, \ref{fig6} and \ref{fig7}.


The next section will give some examples of data analysis concerning the proposed test and compare the results with the $t$ test. We used $14$ artificial data sets which the first $10$ were taken randomly from the internet. The data can be seen in Table \ref{Tab2}.

\subsection{Some examples} \label{sec33}
In this section, some examples data sets are provided and used as an example how to take the decision by using the special case of the cross variance test $T^{*}$ (= the $T$ test when $\sigma_{X}=\sigma_{Y}$). For this $T^{*}$ test, the first step is making sure the variances of the samples are equal. There are some tests for this purpose, in this paper we implement the F.test in R.

The data sets are shown as in Table \ref{Tab2} and the computation results for the F.test are shown in Table \ref{Tab3}. Tables \ref{Tab4} and \ref{Tab5} provide the results of the decision for both methods: the proposed and the $t$ tests. 

In the proposed test, the null hypothesis is rejected if $t^{*}_{o} < t^{*}_{\alpha}$ or if the p-value of the special case of the cross variance test less than $\alpha$, where $t^{*}_{o}$ is the statistics from the observed sample and p-value $=P\left(t^{*}_{o} < t^{*}_{\alpha}\right)$. In the computation we use $\alpha = 0.01$.

From Table \ref{Tab3} we can infer that data set 13 has unequal variances, therefore it will be excluded from further computation.

From the Table \ref{Tab4} we can see that the p-values and decisions from the proposed and the $t$ tests are equal, except for the data set 4. It is the only one data set which has different sample size. When the sample size of two samples is different, we provide two options of $n$: the max$(n_{1},n_{2})$ or the average between $(n_{1},n_{2})$. The example of this is shown in Table \ref{Tab5}.

\section{Summary and Remarks} \label{sec4}

In this paper we have introduced the cross-variance concept, a new test based on the cross variance, the special case of the cross variance (as the proposed test) and the new probability density functions.  A simulation study has been conducted to compute the power and the error type 1 rate of the proposed and the $t$ tests.

The simulation study of the power shows that the proposed and the $t$ tests has the same power. Furthermore the p-value and the error type I rate under the null hypothesis of the proposed and $t$ tests are exactly equal. These results suggest that the proposed test could be used as an alternative test to detect whether there are difference between the means of two independent normal populations, where the variance and sample size is equal. In case the sample size is unequal we proposed to choose $n= \textrm{max}\left(n_{1},n_{2}\right)$ or the average of $\left(n_{1},n_{2}\right)$.

\textbf{Acknowledgements:} \\
This paper is part of the author's PhD dissertation written under the direction of Professor Istv\'an Berkes. We would like to thank \textit{Prof. Alexandre G. Patriota} for the discussions, comments and suggestions. Financial support from the Austrian Science Fund (FWF), Project P24302-N18 is gratefully acknowledged. 

\textbf{Conflict of interest: -}

\section*{References}
\bibliographystyle{plainnat}
\bibliography{emaRef}

\appendix
\newpage{}
\begin{table}[!h]
\begin{center}
\caption{Data sets, their mean and variance} \label{Tab2}
\resizebox{12cm}{6cm}{
\begin{tabular}{c|l|r|r}
\hline
Data	&Data&Mean&Variance	\\ \hline
1	&X=(5,7,5,3,5,3,3,9)&5.000&4.571	\\ 
 	&Y=(8,1,4,6,6,4,1,2) &4.000&6.571	\\ \hline
2	&X=(0.72,0.68,0.69,0.66,0.57,0.66,0.70,0.63,0.71,0.73)&0.675&0.002 \\
	&Y=(0.71,0.83,0.89,0.57,0.68,0.74,0.75,0.67,0.80,0.78) &0.742&0.008	\\ \hline
3	&X=(42,45,40,37,41,41,48,50,45,46)&43.500&15.833 \\
	&Y=(43,51,56,40,32,54,51,55,50,48) &48.000&57.330	\\ \hline
4	&X=(33,31,34,38,32,28)&32.667&11.067 \\
	&Y=(35,42,43,41) &40.250&12.917	\\ \hline
5	&X=(35,40,12,15,21,14,46,10,28,48,16,30,32,48,31,22,
                 12,39,19,25) &27.150&156.450 \\
	&Y=(2,27,38,31,1,19,1,34,3,1,2,3,2,1,2,1,3,29,37,2) &11.950&213.524	\\ \hline
6	&X=(26 21,22,26,19,22,26,25,24,21,23,23,18,29,22)&23.133&8.552	 \\ 
	&Y=(18, 23, 21,20,20,29,20,16,20,26,21,25,17,18,19) &20.867&12.552	 \\ \hline
7	&X=(520,460,500,470)&487.500&758.333	\\ 
	&Y=(230,270,250,280) &257.500&491.667	\\ \hline
8	&X=(3,0,6,7,4,3,2,1,4)&3.333&5.000	\\
	&Y=(5,1,5,7,10,9,7,11,8) &7.000&9.250	\\ \hline
9	&X=(16,20,21,22,23,22,27,25,27,28) &23.100&13.878	\\ 
	&Y=(19,22,24,24,25,25,26,26,28,32) &25.100&11.878	\\ \hline
10	&X=(91,87,99,77,88,91) &88.833&51.367	\\
	&Y=(101,110,103,93,99,104) &101.667&31.867	\\ \hline
11	&X=(10.11,7.36,6.34,11.83,8.61)&8.850&4.761 	\\ 
	&Y=(3.28, 6.52,2.28,6.66,4.55) &4.658&3.760 	\\ \hline
12	&X=(4.79,4.95,2.52,4.98,4.99) &4.446&1.166	\\ 
	&Y=(7.90,7.51,6.62,7.57,7.49) &7.418&0.227	\\ \hline
13	&X=(3.99,3.98,4.03,4.06, 3.84) &3.980&0.007 \\	
	&Y=(6.68,6.25,6.97,5.75,4.01)) &5.932&1.366 \\ \hline
14	&X=(10.16,8.26,16.23,1.44,0.66)&7.350&41.816	\\
	&Y=c(28.06,8.52,25.39,15.45,16.03) &18.690&63.422	\\ \hline
\end{tabular}
}
\end{center}
\end{table}

\begin{table}[!h]
\begin{center}
\caption{F.test decision} \label{Tab3}
\resizebox{6cm}{1.5cm}{
\begin{tabular}{lr|lr}
\hline
Samples	&Decision	&Samples	&Decision\\ \hline
Data 1& equal variance & Data 8& equal variance	\\
Data 2& equal variance & Data 9& equal variance	\\ 
Data 3& equal variance & Data 10& equal variance	\\
Data 4& equal variance & Data 11& equal variance 	\\ 
Data 5& equal variance & Data 12& equal variance	\\
Data 6& equal variance & Data 13& not equal variance \\	
Data 7& equal variance & Data 14& equal variance	\\ \hline
\end{tabular}
}
\end{center}
\end{table}
\newpage{}
\begin{table}[!h]
\begin{center}
\caption{P-values and decision from the proposed and $t$ tests} \label{Tab4}
\resizebox{6cm}{2.5cm}{
\begin{tabular}{lrrrr}
\hline
\multirow{2}{*}{Sample}	&  \multicolumn{2}{c}{$t$ test} & \multicolumn{2}{c}{proposed test}  \\ \hhline{~----}
	    &p-value &Decision &p-value  &  Decision \\ \hline
Data 1  &0.411    &Accept    &0.411       & Accept    \\ 
Data 2  &0.054    &Accept    &0.054        & Accept     \\ 
Data 3  &0.114    &Accept    &0.113        & Accept    \\ 
Data 4  &0.009    &Reject    &      &   \\ 
Data 5  &0.001    &Reject    &0.001	    & Reject   \\ 
Data 6  &0.067    &Accept   &0.066	    & Accept  \\ 
Data 7  &0.000    &Reject   &0.000	    & Reject   \\ 
Data 8  &0.010    &Accept  &0.010	    & Accept    \\ 
Data 9  &0.229    &Accept  &0.229	    & Accept  \\ 
Data 10& 0.006   &Reject   &0.006	    & Reject  \\ 
Data 11& 0.012   &Accept  & 0.012	   &Accept	 \\ 
Data 12& 0.000   &Reject   &0.000	   & Reject   \\ 
Data 14& 0.039  &Accept   &0.039     & Accept  \\ \hline
\end{tabular}
}
\end{center}
\end{table}
\begin{table}[!h]
\begin{center}
\caption{P-values and decision from the proposed test, Data set 4} \label{Tab5}
\resizebox{3cm}{0.75cm}{
\begin{tabular}{lrr}
\hline
\multirow{2}{*}{Choice of $n$}	&  \multicolumn{2}{c}{Least Square}  \\ \hhline{~--}
	    &p-value  &  Decision 	\\ \hline
Min  &0.021       & Accept    \\ 
Max  &0.004        & Reject     \\ 
Average  &0.009        & Reject    \\ \hline
\end{tabular}
}
\end{center}
\end{table}

\end{document}